\documentclass[12pt]{article}
\usepackage{eqsection,subeqnarray,indent,amsfonts,amssymb}

\begin{document}

\title{Chromatic Roots are Dense \\ in the Whole Complex Plane}

\author{
  \\[-1mm]
  {Alan D. Sokal}                   \\
  {\it Department of Physics}       \\
  {\it New York University}         \\
  {\it 4 Washington Place}          \\
  {\it New York, NY 10003 USA}      \\
  {\tt SOKAL@NYU.EDU}          \\
  \\
}
\vspace{0.5cm}

\date{December 18, 2000 \\[1.5mm] revised December 28, 2000 \\[1.5mm]
      final version August 12, 2003 \\[2mm]
      to appear in {\em Combinatorics, Probability and Computing}\/
     }
\maketitle
\thispagestyle{empty}   % Suppress page number on front page.

\begin{abstract}
I show that the zeros of the chromatic polynomials $P_G(q)$
for the generalized theta graphs $\Theta^{(s,p)}$
are, taken together, dense in the whole complex plane
with the possible exception of the disc $|q-1| < 1$.
The same holds for their dichromatic polynomials
(alias Tutte polynomials, alias Potts-model partition functions)
$Z_G(q,v)$ outside the disc $|q+v| < |v|$.
An immediate corollary is that the chromatic roots of
not-necessarily-planar graphs are dense in the whole complex plane.
The main technical tool in the proof of these results
is the Beraha--Kahane--Weiss theorem
on the limit sets of zeros for certain sequences of analytic functions,
for which I give a new and simpler proof.
\end{abstract}

\vspace{0.5cm}

\clearpage

\newcommand{\be}{\begin{equation}}
\newcommand{\ee}{\end{equation}}
\newcommand{\<}{\langle}
\renewcommand{\>}{\rangle}
\newcommand{\widebar}{\overline}
\def\reff#1{(\protect\ref{#1})}
\def\spose#1{\hbox to 0pt{#1\hss}}
\def\ltapprox{\mathrel{\spose{\lower 3pt\hbox{$\mathchar"218$}}
 \raise 2.0pt\hbox{$\mathchar"13C$}}}
\def\gtapprox{\mathrel{\spose{\lower 3pt\hbox{$\mathchar"218$}}
 \raise 2.0pt\hbox{$\mathchar"13E$}}}
\def\textprime{${}^\prime$}
\def\proof{\par\medskip\noindent{\sc Proof.\ }}
\def\qed{\hbox{\hskip 6pt\vrule width6pt height7pt depth1pt \hskip1pt}\bigskip}
\def\proofof#1{\bigskip\noindent{\sc Proof of #1.\ }}
\def\half{ {1 \over 2} }
\def\third{ {1 \over 3} }
\def\twothird{ {2 \over 3} }
\def\smfrac#1#2{\textstyle{#1\over #2}}
\def\smhalf{ \smfrac{1}{2} }
\newcommand{\real}{\mathop{\rm Re}\nolimits}
\renewcommand{\Re}{\mathop{\rm Re}\nolimits}
\newcommand{\imag}{\mathop{\rm Im}\nolimits}
\renewcommand{\Im}{\mathop{\rm Im}\nolimits}
\newcommand{\sgn}{\mathop{\rm sgn}\nolimits}
\def\hboxscript#1{ {\hbox{\scriptsize\it #1}} }
\def\Sfin{ S_{\hboxscript{fin}} }
\def\betatilde{{\widetilde{\beta}}}

\newcommand{\restrict}{\upharpoonright}
 
\def\scra{\mathcal{A}}
\def\scrc{\mathcal{C}}
\def\scrf{\mathcal{F}}
\def\scrg{\mathcal{G}}
\def\scrh{\mathcal{H}}
\def\scrl{\mathcal{L}}
\def\scro{\mathcal{O}}
\def\scrp{\mathcal{P}}
\def\scrr{\mathcal{R}}
\def\scrs{\mathcal{S}}
\def\scrt{\mathcal{T}}
\def\scru{\mathcal{U}}
\def\scrv{\mathcal{V}}
\def\scrz{\mathcal{Z}}

\def\q{{\sf q}}

\def\Z{{\mathbb Z}}
\def\R{{\mathbb R}}
\def\C{{\mathbb C}}
\def\N{{\mathbb N}}

\def\veff{v_{\hboxscript{eff}}}
\def\qve{{q, \{v_e\}}}
\def\zgxconny{Z_G^{(x \leftrightarrow y)}}
\def\zgxnoconny{Z_G^{(x \not\leftrightarrow y)}}
 
\newtheorem{theorem}{Theorem}[section]
\newtheorem{proposition}[theorem]{Proposition}
\newtheorem{lemma}[theorem]{Lemma}
\newtheorem{corollary}[theorem]{Corollary}
\newtheorem{definition}[theorem]{Definition}
\newtheorem{question}[theorem]{Question}

% Array for subscripts

\newenvironment{sarray}{
          \textfont0=\scriptfont0
          \scriptfont0=\scriptscriptfont0
          \textfont1=\scriptfont1
          \scriptfont1=\scriptscriptfont1
          \textfont2=\scriptfont2
          \scriptfont2=\scriptscriptfont2
          \textfont3=\scriptfont3
          \scriptfont3=\scriptscriptfont3
        \renewcommand{\arraystretch}{0.7}
        \begin{array}{l}}{\end{array}}
 
\newenvironment{scarray}{
          \textfont0=\scriptfont0
          \scriptfont0=\scriptscriptfont0
          \textfont1=\scriptfont1
          \scriptfont1=\scriptscriptfont1
          \textfont2=\scriptfont2
          \scriptfont2=\scriptscriptfont2
          \textfont3=\scriptfont3
          \scriptfont3=\scriptscriptfont3
        \renewcommand{\arraystretch}{0.7}
        \begin{array}{c}}{\end{array}}

\section{Introduction}   \label{sec1}

\subsection{(Di)chromatic polynomials and the Potts model}  \label{sec1.1}

The polynomials studied in this paper arise independently
in graph theory and in statistical mechanics.
It is appropriate, therefore, to begin by explaining each of these contexts.
Specialists in these fields are warned that they will find at least one
(and perhaps both) of these summaries excruciatingly boring;
they can skip them.

Let $G = (V,E)$ be a finite undirected graph\footnote{
   In this paper a ``graph'' is allowed to have
   loops and/or multiple edges unless explicitly stated otherwise.
}
with vertex set $V$ and edge set $E$.
For each positive integer $q$,
let $P_G(q)$ be the number of ways that the vertices of $G$
can be assigned ``colors'' from the set $\{ 1,2,\ldots,q \}$
in such a way that adjacent vertices always receive different colors.
It is not hard to show (see below) that $P_G(q)$ is the restriction
to $\Z_+$ of a polynomial in $q$.
This (obviously unique) polynomial is called the
{\bf chromatic polynomial}\/ of $G$,
and can be taken as the {\em definition}\/ of $P_G(q)$ for
arbitrary real or complex values of $q$.\footnote{
   See \cite{Read_68,Read_88} for excellent reviews on
   chromatic polynomials, and \cite{Chia_97} for an extensive bibliography.
}

The chromatic polynomial was introduced in 1912 by Birkhoff \cite{Birkhoff_12}.
The original hope was that study of the real or complex zeros of $P_G(q)$
might lead to an analytic proof of the Four-Color Conjecture
\cite{Ore_67,Saaty_77},
which states that $P_G(4) > 0$ for all loopless planar graphs $G$.
To date this hope has not been realized, although combinatoric proofs
of the Four-Color Theorem have been found
\cite{Appel_77a,Appel_77b,Appel_89,Robertson_97,Thomas_98}.
Even so, the zeros of $P_G(q)$ are interesting in their own right
and have been extensively studied.
Most of the available theorems concern real zeros
\cite{Birkhoff_46,Tutte_70a,Woodall_77,Woodall_92a,Woodall_92b,%
Jackson_93,Woodall_97,Thomassen_97,Edwards_98,Thomassen_00},
but there have been some theorems on complex zeros
\cite{Biggs_72,Beraha_79,Beraha_80,Wakelin_92,Brenti_94,Brown_98a,Brown_98b,%
Brown-Hickman_99a,Brown-Hickman_99b,Sokal_00,Sokal_00b,gen_theta}
as well as numerical studies and computations for special families
\cite{Hall_65,Berman_69,Biggs_72,Farrell_80,Baxter_86,Baxter_87,Read_91,%
Shrock_97a,Shrock_97b,Shrock_97c,Shrock_97d,Shrock_98a,Shrock_98b,%
Shrock_98c,Tsai_98,Shrock_98e,Shrock_99a,Shrock_99b,Shrock_99c,Shrock_99d,%
Shrock_99f,Shrock_99g,Biggs_99c,%
Chang_00c,Chang_00e,Chang_00a,Chang_00b,Chang_00f,Chang_00d,Chang_00h,%
Chang_00i,Chang_00g,Biggs_99b,Biggs_99a,Shrock_00a,Shrock_00b,%
Salas-Sokal_transfer1,Jacobsen-Salas_transfer2,Jacobsen-Salas-Sokal_transfer3}.

A more general polynomial can be obtained as follows:
Assign to each edge $e \in E$ a real or complex weight $v_e$.
Then define
\be
   Z_G(q, \{v_e\})   \;=\;
   \sum_{ \sigma }  \,  \prod_{e \in E}  \,
      \biggl[ 1 + v_e \delta(\sigma(x_1(e)), \sigma(x_2(e))) \biggr]
   \;,
 \label{eq1.1}
\ee
where the sum runs over all maps $\sigma\colon\, V \to \{ 1,2,\ldots,q \}$,
the $\delta$ is the Kronecker delta,
and $x_1(e), x_2(e) \in V$ are the two endpoints of the edge $e$
(in arbitrary order).
It is not hard to show (see below)
that $Z_G(q, \{v_e\})$ is the restriction to $q \in \Z_+$ of
a polynomial in $q$ and $\{v_e\}$.
If we take $v_e = -1$ for all $e$, this reduces to the chromatic polynomial.
If we take $v_e = v$ for all $e$, this defines a two-variable polynomial
$Z_G(q,v)$ that was introduced
implicitly by Whitney \cite{Whitney_32a,Whitney_32b,Whitney_33}
and explicitly by Tutte \cite{Tutte_47,Tutte_54};
it is known variously (modulo trivial changes of variable)
as the {\bf dichromatic polynomial}\/, the {\bf dichromate}\/,
the {\bf Whitney rank function}\/ or the {\bf Tutte polynomial}\/
\cite{Welsh_93,Biggs_93}.\footnote{
   The Tutte polynomial $T_G(x,y)$ is conventionally defined as
   \protect\cite[p.~45]{Welsh_93} \protect\cite[pp.~73, 101]{Biggs_93}
   $$
     T_G(x,y)   \;=\;
     \sum\limits_{E' \subseteq E} (x-1)^{k(E')-k(E)} \,
                                (y-1)^{|E'| + k(E') - |V|}
   $$
   where $k(E')$ is the number of connected components
   in the subgraph $(V,E')$.
   Comparison with (\protect\ref{eq1.2}) below yields
   $$
     T_G(x,y)   \;=\;   (x-1)^{-k(E)} \, (y-1)^{-|V|} \,
                        Z_G \bigl( (x-1)(y-1), \, y-1 \bigr)   \;.
   $$
}

In statistical mechanics, \reff{eq1.1} is known as the
partition function of the
{\bf \mbox{\boldmath $q$}-state Potts model}\/.
%% \footnote{
%%    The Potts model \cite{Potts_52} was invented in the early 1950's by Domb
%%    (see \cite{Domb_74}).
%%    The $q=2$ case, known as the Ising model \cite{Ising_25},
%%    was invented in 1920 by Lenz \cite{Lenz_20}
%%    (see \cite{Brush_67,Kobe_97,Ising_obit}).
%%    The $q=4$ case, which is a special case of the Ashkin--Teller model,
%%    was invented in 1943 by Ashkin and Teller \cite{Ashkin-Teller_43}.
%% }
In the Potts model \cite{Potts_52,Wu_82,Wu_84},
an ``atom'' (or ``spin'') at site $x \in V$ can exist in any one of
$q$ different states (where $q$ is an integer $\ge 1$).
The {\bf energy}\/ of a configuration is the sum, over all edges $e \in E$,
of $0$ if the spins at the two endpoints of that edge are unequal
and $-J_e$ if they are equal.
The {\bf Boltzmann weight}\/ of a configuration is then $e^{-\beta H}$,
where $H$ is the energy of the configuration
and $\beta \ge 0$ is the inverse temperature.
The {\bf partition function}\/ is the sum, over all configurations,
of their Boltzmann weights.
Clearly this is just a rephrasing of \reff{eq1.1},
with $v_e = e^{\beta J_e} - 1$.
A coupling $J_e$ (or $v_e$)
is called {\bf ferromagnetic}\/ if $J_e \ge 0$ ($v_e \ge 0$)
and {\bf antiferromagnetic}\/ if $-\infty \le J_e \le 0$ ($-1 \le v_e \le 0$).
%%  and {\bf unphysical}\/ if $v_e < -1$.

To see that $Z_G(q, \{v_e\})$ is indeed a polynomial in its arguments
(with coefficients that are in fact 0 or 1), we proceed as follows:
In \reff{eq1.1}, expand out the product over $e \in E$,
and let $E' \subseteq E$ be the set of edges for which the term
$v_e \delta(\sigma(x_1(e)), \sigma(x_2(e)))$ is taken.
Now perform the sum over configurations $\sigma$:
in each connected component of the subgraph $(V,E')$
the spin value $\sigma(x)$ must be constant,
and there are no other constraints.
Therefore,
\be
   Z_G(q, \{v_e\})   \;=\;
   \sum_{ E' \subseteq E }  q^{k(E')}  \prod_{e \in E'}  v_e
   \;,
 \label{eq1.2}
\ee
where $k(E')$ is the number of connected components
(including isolated vertices) in the subgraph $(V,E')$.
The expansion \reff{eq1.2} was discovered by
Birkhoff \cite{Birkhoff_12} and Whitney \cite{Whitney_32a}
for the special case $v_e = -1$ (see also Tutte \cite{Tutte_47,Tutte_54});
in its general form it is due to
Fortuin and Kasteleyn \cite{Kasteleyn_69,Fortuin_72}
(see also \cite{Edwards-Sokal}).
We take \reff{eq1.2} as the {\em definition}\/ of $Z_G(q, \{v_e\})$
for arbitrary complex $q$ and $\{v_e\}$.

In statistical mechanics, a very important role is played by the
complex zeros of the partition function.
This arises as follows \cite{Yang-Lee_52}:
Statistical physicists are interested in {\em phase transitions}\/,
namely in points where one or more physical quantities
(e.g.\ the energy or the magnetization)
depend nonanalytically (in many cases even discontinuously)
on one or more control parameters
(e.g.\ the temperature or the magnetic field).
Now, such nonanalyticity is manifestly impossible in \reff{eq1.1}/\reff{eq1.2}
for any finite graph $G$.
Rather, phase transitions arise only in the {\em infinite-volume limit}\/.
That is, we consider some countably infinite graph
$G_\infty = (V_\infty, E_\infty)$
--- usually a regular lattice, such as $\Z^d$ with nearest-neighbor edges ---
and an increasing sequence of finite subgraphs $G_n = (V_n, E_n)$.
It can then be shown
(under modest hypotheses on the $G_n$)
%%%  \cite[Section I.2]{Israel_79})
that the {\bf (limiting) free energy per unit volume}\/
\be
   f_{G_\infty}(q,v)   \;=\;
   \lim_{n \to \infty}   |V_n|^{-1}  \log Z_{G_n}(q,v)
 \label{limiting_free_energy}
\ee
exists for all {\em nondegenerate physical}\/ values
of the parameters\footnote{
   Here ``physical'' means that the weights are nonnegative,
   so that the model has a probabilistic interpretation;
   and ``nondegenerate'' means that we exclude the limiting cases
   $v=-1$ in (a) and $q=0$ in (b), which cause difficulties
   due to the existence of configurations having zero weight.
},
namely either
\begin{quote}
\begin{itemize}
  \item[(a)]     $q$ integer $\ge 1$ and $-1 < v < \infty$
     \quad  [using \reff{eq1.1}:  see e.g.\ \cite[Section I.2]{Israel_79}]
  \item[or (b)]  $q$ real $> 0$ and $0 \le v < \infty$
     \quad  [using \reff{eq1.2}:  see \cite[Theorem 4.1]{Grimmett_95}
                                  and \cite{Grimmett_78,Seppalainen_98}].
\end{itemize}
\end{quote}
This limit $f_{G_\infty}(q,v)$ is in general a continuous function of $v$;
but it can fail to be a real-analytic function of $v$,
because complex singularities of $\log Z_{G_n}(q,v)$
--- namely, complex zeros of $Z_{G_n}(q,v)$ ---
can approach the real axis in the limit $n \to\infty$.
Therefore, the possible points of physical phase transitions
are precisely the real limit points of such complex zeros
(see Theorem~\ref{thm3.1}).
As a result, theorems that constrain the possible location of
complex zeros of the partition function are of great interest.
In particular, theorems guaranteeing that a certain complex domain
is free of zeros are often known as {\em Lee-Yang theorems}\/.\footnote{
   The first such theorem, concerning the behavior of the ferromagnetic
   Ising model at complex magnetic field, was proven by Lee and Yang
   \cite{Lee-Yang_52} in 1952.
   A partial bibliography (through 1980) of generalizations of this result
   can be found in \cite{Lieb-Sokal_81}.
}

In summary, both graph theorists and statistical mechanicians are interested
in the zeros of $Z_G(q,v)$.
Graph theorists most often fix $v = -1$ and look for zeros in the
complex $q$-plane,
while statistical mechanicians most often fix $q$ real $> 0$
(usually but not always an integer)
and look for zeros in the complex $v$-plane.
But these inclinations are not hard-and-fast;
both groups have seen the value of investigating the more general
two-variable problem.
The analysis in this paper will, in fact, be an illustration of
the value of doing so,
even if one is ultimately interested in a specific one-variable
specialization of $Z_G(q,v)$.

\subsection{A Lee-Yang theorem for chromatic polynomials?}  \label{sec1.2}

Let me now review some known facts about the {\em real}\/ zeros of the
chromatic polynomial $P_G(q)$, in order to motivate some conjectures
concerning the {\em complex}\/ zeros:

1)  It is not hard to show that for any loopless graph $G$ with $n$ vertices,
$(-1)^n P_G(q) > 0$ for real $q < 0$ \cite{Read_88}.
It is then natural to ask whether the absence of negative real zeros
might be the tip of the iceberg of a Lee-Yang theorem:
that is, might there exist a {\em complex}\/ domain $D$
containing $(-\infty,0)$ that is zero-free for all $P_G$?
One's first guess is that the half-plane $\Re q < 0$ might be zero-free
\cite{Farrell_80}.
This turns out to be false:
for about a decade, examples have been known of loopless graphs $G$
that have chromatic roots with slightly negative real part
\cite{Baxter_87,Read_91,Shrock_98b,Shrock_99b,Shrock_00a,%
Brown-Hickman_99a,Brown-Hickman_99b,Salas-Sokal_transfer1};
and examples have very recently been constructed with
arbitrarily negative real part \cite[equation (3.12)]{Shrock_98e}.
Nevertheless, it is not ruled out that some smaller domain
$D \supset (-\infty,0)$ might be zero-free.

2) For any loopless {\em planar}\/ graph $G$,
Birkhoff and Lewis \cite{Birkhoff_46} proved in 1946 that
$P_G(q) > 0$ for real $q \ge 5$;\footnote{
   See also \cite[Theorem 1]{Woodall_97}
   and \cite[Theorem 3.1 ff.]{Thomassen_97}
   for alternative proofs of a more general result.
}
we now know that $P_G(4) > 0$
\cite{Appel_89,Robertson_97};
and it is very likely true (though not yet proven as far as I know)
that $P_G(q) > 0$ also for $4 < q < 5$.
Thus, it is natural to conjecture that might exist
a {\em complex}\/ domain $D$ containing $(4,\infty)$ [or $(5,\infty)$]
that is zero-free for all planar $P_G$.
One's first guess might be that $\Re q > 4$ works.
This again turns out to be false:
examples are known of loopless planar graphs $G$
that have chromatic roots with real part as large as $\approx 4.2$
\cite{Baxter_87,Jacobsen-Salas-Sokal_transfer3}.
Nevertheless, it is not ruled out that some smaller domain
$D \supset (4,\infty)$ might be zero-free.

As with most of my conjectures, these two are false;
but what is interesting is that they are {\em utterly, spectacularly false}\/,
for I can prove:
\begin{theorem}
   \label{thm1.1}
There is a countably infinite family of planar (in fact, series-parallel)
graphs whose chromatic roots are, taken together,
dense in the entire complex $q$-plane
with the possible exception of the disc $|q-1| < 1$.
\end{theorem}
As far as I know, it was until now an open question whether
the closure of the set of all chromatic roots
(of all graphs taken together) even has
nonzero two-dimensional Lebesgue measure.
Theorem~\ref{thm1.1} answers this question in a most spectacular way.

The graphs arising in Theorem~\ref{thm1.1} are
``generalized theta graphs'' $\Theta^{(s,p)}$
obtained by parallel-connecting $p$ chains each of which has
$s$ edges in series.\footnote{
   More generally,
   the generalized theta graph $\Theta_{s_1,\ldots,s_p}$ consists of
   endvertices $x,y$ connected by $p$ internally disjoint paths
   of lengths $s_1, \ldots, s_p \ge 1$ \cite{gen_theta}.
   The graphs arising in Theorem~\ref{thm1.1}
   thus correspond to the special case $s_1 = \ldots = s_p = s$.
   See Section~\ref{sec2.3} below for a computation of the
   Potts-model partition function for an arbitrary $\Theta_{s_1,\ldots,s_p}$.
}
Theorem~\ref{thm1.1} is in fact a corollary of a more general result
for the two-variable polynomials $Z_G(q,v)$:
\begin{theorem}
   \label{thm1.2}
Fix complex numbers $q_0,v_0$ satisfying $|v_0| \le |q_0 + v_0|$.
Then, for each $\epsilon > 0$, there exist $s_0 \in \N$
and a map $p_0 \colon\, \N \cap [s_0,\infty) \to \N$
such that for all $s \ge s_0$ and $p \ge p_0(s)$:
\begin{itemize}
   \item[(a)]  If $v_0 \neq 0$, then
      $Z_{\Theta^{(s,p)}}(\,\cdot\,, v_0)$ has a zero in the disc
      $|q - q_0| < \epsilon$.
   \item[(b)]  $Z_{\Theta^{(s,p)}}(q_0, \,\cdot\,)$ has a zero in the disc
      $|v - v_0| < \epsilon$.
\end{itemize}
\end{theorem}
(Setting $v_0 = -1$, Theorem~\ref{thm1.2}(a) implies Theorem~\ref{thm1.1}.)

As Jason Brown pointed out to me (why didn't I notice it myself?),
an immediate corollary of Theorem~\ref{thm1.1} is:
\begin{corollary}
   \label{cor1.3}
There is a countably infinite family of (not-necessarily-planar)
graphs whose chromatic roots are, taken together,
dense in the entire complex $q$-plane.
\end{corollary}
Indeed, it suffices to consider the union of the two families
$\Theta^{(s,p)}$ and $\Theta^{(s,p)} + K_2$
(the latter is the join of $\Theta^{(s,p)}$
 with the complete graph on two vertices)
and to recall that $P_{G + K_n}(q) = q(q-1) \cdots (q-n+1) \, P_G(q-n)$.

It is an open question whether the chromatic roots of planar
(or perhaps even series-parallel) graphs are dense in the disc $|q-1| < 1$.
I have some partial results on this question,
but since they are not yet definitive,
I shall report them elsewhere.

The methods of this paper actually prove a result stronger than
Theorem~\ref{thm1.2}, namely:

\begin{theorem}
   \label{thm1.2bis}
\begin{itemize}
   \item[(a)]  Fix a complex number $v_0 \neq 0$.
      Then, for each $\epsilon > 0$ and $R < \infty$,
      there exist $s_0 \in \N$
      and a map $p_0 \colon\, \N \cap [s_0,\infty) \to \N$
      such that for all $s \ge s_0$ and $p \ge p_0(s)$,
      the zeros of $Z_{\Theta^{(s,p)}}(\,\cdot\,, v_0)$
      come within $\epsilon$ of every point in the region
      $\{ q \in \C \colon\; |q + v_0| \ge |v_0| \hbox{ and } |q| \le R\}$.
   \item[(b)]  Fix a complex number $q_0$.
      Then, for each $\epsilon > 0$ and $R < \infty$,
      there exist $s_0 \in \N$
      and a map $p_0 \colon\, \N \cap [s_0,\infty) \to \N$
      such that for all $s \ge s_0$ and $p \ge p_0(s)$,
      the zeros of $Z_{\Theta^{(s,p)}}(q_0, \,\cdot\,)$
      come within $\epsilon$ of every point in the region
      $\{ v \in \C \colon\; |v| \le |q_0 + v| \hbox{ and } |v| \le R\}$.
\end{itemize}
\end{theorem}
I thank Roberto Fern\'andez for posing the question of whether
something like Theorem~\ref{thm1.2bis} might be true.

\subsection{Sketch of the proof}  \label{sec1.3}

The intuition behind Theorem~\ref{thm1.2} is based on recalling
the rules for parallel and series combination of Potts edges
(see Section \ref{sec2} for details):
\begin{itemize}
   \item[\quad] Parallel:
      $v_{\hboxscript{eff}} = v_1 + v_2 + v_1 v_2$
      \qquad\,\,\, (mnemonic: $1 + v$ multiplies)
   \item[\quad] Series:
      $v_{\hboxscript{eff}} = v_1 v_2 / (q + v_1 + v_2)$
      \qquad (mnemonic: $v/(q+v)$ multiplies)
\end{itemize}
In particular, if $0 < |v/(q+v)| < 1$, then putting a large number $s$ of edges
in series drives the effective coupling $v_{\hboxscript{eff}}$
to a small (but nonzero) number;
moreover, by small perturbations of $v$ and/or $q$
we can give $v_{\hboxscript{eff}}$ any phase we please.
But then, by putting a large number $p$ of such chains in parallel,
we can make the resulting $v_{\hboxscript{eff}}$ lie anywhere
in the complex plane we please.
In particular, we can make $v_{\hboxscript{eff}}$ equal to $-q$,
which causes the partition function $Z_{\Theta^{(s,p)}}$ to be zero.

To convert this intuition into a proof,
I employ a complex-variables result due to
Beraha, Kahane and Weiss \cite{BKW_75,BKW_78,Beraha_79,Beraha_80},
which I slightly generalize as follows:
Let $D$ be a domain (connected open set) in $\C$,
and let $\alpha_1,\ldots,\alpha_m,\beta_1,\ldots,\beta_m$ ($m \ge 2$)
be analytic functions on $D$, none of which is identically zero.
For each integer $n \ge 0$, define
\be
   f_n(z)   \;=\;   \sum\limits_{k=1}^m \alpha_k(z) \, \beta_k(z)^n
   \;.
   \label{def_fn}
\ee
We are interested in the zero sets
\be
   \scrz(f_n)   \;=\;   \{z \in D \colon\;  f_n(z) = 0 \}
\ee
and in particular in their limit sets as $n\to\infty$:
\begin{eqnarray}
   \liminf \scrz(f_n)   & = &  \{z \in D \colon\;
   \hbox{every neighborhood $U \ni z$ has a nonempty intersection} \nonumber\\
      & & \qquad \hbox{with all but finitely many of the sets } \scrz(f_n) \}
   \\[4mm]
   \limsup \scrz(f_n)   & = &  \{z \in D \colon\;
   \hbox{every neighborhood $U \ni z$ has a nonempty intersection} \nonumber\\
      & & \qquad \hbox{with infinitely many of the sets } \scrz(f_n) \}
\end{eqnarray}
Let us call an index $k$ {\em dominant at $z$}\/ if
$|\beta_k(z)| \ge |\beta_l(z)|$ for all $l$ ($1 \le l \le m$);
and let us write
\be
   D_k  \;=\;  \{ z \in D \colon\;  k \hbox{ is dominant at } z  \}
   \;.
\ee
Then the limiting zero sets can be completely characterized as follows:

\begin{theorem}
   \label{thm1.5}
Let $D$ be a domain in $\C$,
and let $\alpha_1,\ldots,\alpha_m,\beta_1,\ldots,\beta_m$ ($m \ge 2$)
be analytic functions on $D$, none of which is identically zero.
Let us further assume a ``no-degenerate-dominance'' condition:
there do not exist indices $k \neq k'$
such that $\beta_k \equiv \omega \beta_{k'}$ for some constant $\omega$
with $|\omega| = 1$ and such that $D_k$ ($= D_{k'}$)
has nonempty interior.
For each integer $n \ge 0$, define $f_n$ by
$$
   f_n(z)   \;=\;   \sum\limits_{k=1}^m \alpha_k(z) \, \beta_k(z)^n
   \;.
$$
Then $\liminf \scrz(f_n) = \limsup \scrz(f_n)$,
and a point $z$ lies in this set if and only if either
\begin{itemize}
   \item[(a)]  There is a unique dominant index $k$ at $z$,
       and $\alpha_k(z) =0$;  or
   \item[(b)]  There are two or more dominant indices at $z$.
\end{itemize}
\end{theorem}
Note that case (a) consists of isolated points in $D$,
while case (b) consists of curves
(plus possibly isolated points where all the $\beta_k$ vanish simultaneously).
Beraha--Kahane--Weiss considered the special case of Theorem~\ref{thm1.5}
in which the $f_n$ are polynomials satisfying a linear finite-order
recurrence relation, and they assumed a slightly stronger nondegeneracy
condition.  Their theorem is all we really need to prove
Theorems~\ref{thm1.2} and \ref{thm1.2bis},
but the general result is more natural
and its proof is no more difficult.
Indeed, my proof (see Section~\ref{sec3}) is quite a bit simpler
than the original proof of Beraha--Kahane--Weiss \cite{BKW_78}
(though also less powerful in that it gives no information on
 the {\em rate}\/ of convergence of the zeros of $f_n$
 to their limiting set).

The next step is to notice that the dichromatic polynomial for
the generalized theta graph $\Theta^{(s,p)}$
has precisely the form \reff{def_fn} with $m=2$:
\be
   Z_{\Theta^{(s,p)}}(q,v)   \;=\;
   {[(q+v)^s + (q-1)v^s]^p \,+\, (q-1) [(q+v)^s - v^s]^p
    \over
    q^{p-1}
   }
 \label{Z_gen_theta}
\ee
(see Section~\ref{sec2} for the easy calculation).
It follows from Theorem \ref{thm1.5} that when $p\to\infty$ at fixed $s$,
the zeros of $Z_{\Theta^{(s,p)}}$ accumulate where\footnote{
   For Theorems~\ref{thm1.2}(a) and \ref{thm1.2bis}(a),
   the no-degenerate-dominance condition
   of Theorem \ref{thm1.5} is satisfied whenever $v_0 \neq 0$.
   For Theorems~\ref{thm1.2}(b) and \ref{thm1.2bis}(b),
   the no-degenerate-dominance condition
   is satisfied whenever $q_0 \neq 0$;  but if $q_0 = 0$, then
   $Z_{\Theta^{(s,p)}}(q_0, \,\cdot\,)$ is identically zero
   and the assertion is trivially true.
}
\be
   \left| 1 \,+\, (q-1) \left({v \over q+v}\right) ^{\! s} \right|
   \;=\;
   \left| 1 \,-\, \left({v \over q+v}\right) ^{\! s} \right|
   \;.
  \label{fixed_s_curve}
\ee
I then use the following lemma to handle the limit $s \to\infty$:

\begin{lemma}
   \label{lemma1.6}
Let $F_1,F_2,G$ be analytic functions on a disc $|z| < R$
satisfying $|G(0)| \le 1$ and $G \not\equiv \hbox{constant}$.
Then, for each $\epsilon > 0$, there exists $s_0 < \infty$
such that for all integers $s \ge s_0$ the equation
\be
   |1 + F_1(z) G(z)^s|   \;=\;  |1 + F_2(z) G(z)^s|
\ee
has a solution in the disc $|z| < \epsilon$.
\end{lemma}
Theorems~\ref{thm1.2} and \ref{thm1.2bis} are an almost immediate consequence
(see Section~\ref{sec5} for details).

\subsection{Plan of this paper}  \label{sec1.4}

The plan of this paper is as follows:
In Section~\ref{sec2} I discuss some identities satisfied by
the Potts-model partition function $Z_G(q, \{v_e\})$;
in particular, I show what happens when a 2-rooted graph $(G,x,y)$
is inserted in place of an edge $e_*$ in some other graph $H$.
As a special case, I obtain the well-known rules
for series and parallel combination of Potts edges,
which allow one to compute in a straightforward way
the Potts-model partition function $Z_G(q, \{v_e\})$
for any series-parallel graph $G$.
Specializing further, I derive the formula \reff{Z_gen_theta} for the
dichromatic polynomial $Z_G(q,v)$ of the generalized theta graphs
$\Theta^{(s,p)}$.
In Section~\ref{sec3} I prove some theorems ---
which seem to me of considerable interest in their own right ---
concerning the limit sets of zeros for certain sequences of analytic functions.
As a corollary, I obtain a simple proof of Theorem~\ref{thm1.5}.
In Section~\ref{sec4} I prove a strengthened version of Lemma~\ref{lemma1.6}.
In Section~\ref{sec5} I complete the proof of
Theorems~\ref{thm1.2} and \ref{thm1.2bis}.
In Section~\ref{sec6} I digress to study the {\em real}\/ chromatic roots
of the graphs $\Theta^{(s,p)}$.
In Section~\ref{sec7} I discuss some variants of the construction
employed in this paper.
In Section~\ref{sec8} I discuss some open questions.

The two appendices provide further examples of the power
of the identities derived in Section~\ref{sec2}.
In Appendix~\ref{app.A} I give a simple proof of the
Brown--Hickman \cite{Brown-Hickman_99b} theorem on chromatic roots
of large subdivisions.
In Appendix~\ref{app.B} I extend Thomassen's \cite{Thomassen_97}
construction concerning the chromatic roots of 2-degenerate graphs.

\section{Some Identities for Potts Models}   \label{sec2}

In this section I discuss some identities for Potts-model partition functions
and use them to calculate $Z_G(\qve)$ for generalized theta graphs.
There are two alternative approaches to proving such identities:
one is to prove the identity directly for complex $q$,
using the Fortuin--Kasteleyn representation \reff{eq1.2};
the other is to prove the identity first for {\em positive integer}\/ $q$,
using the spin representation \reff{eq1.1},
and then to extend it to complex $q$ by arguing that two polynomials
(or rational functions) that coincide at infinitely many points must be equal.
The latter approach is perhaps less elegant,
but it is often simpler or more intuitive.

\subsection{Restricted Potts-model partition functions for 2-rooted graphs} \label{sec2.1}

Let $G=(V,E)$ be a finite graph, and let $x,y$ be distinct vertices of $G$.
We define $G \bullet xy$ to be the graph in which $x$ and $y$ are contracted
to a single vertex.  (N.B.:  If $G$ contains one or more edges $xy$,
then these edges are {\em not}\/ deleted, but become loops in $G \bullet xy$.)
There is a canonical one-to-one correspondence between the edges of $G$
and the edges of $G \bullet xy$;  for simplicity (though by slight abuse of notation)
we denote an edge of $G$ and the corresponding edge of $G \bullet xy$
by the same letter.  In particular, we can apply a given set of edge weights
$\{ v_e \} _{e \in E}$ to both $G$ and $G \bullet xy$.

Let us now define
\begin{eqnarray}
   \zgxconny(\qve)   & = & \!\!\!
   \sum_{\begin{scarray}
            E' \subseteq E \\
            \hboxscript{$E'$ connects $x$ to $y$}
         \end{scarray}
        }
   \!\!\!\!\!  q^{k(E')} \;  \prod_{e \in E'}  v_e
   \\[4mm]
   \zgxnoconny(\qve)   & = &
   \!\!\!
   \sum_{\begin{scarray}
            E' \subseteq E \\
            \hboxscript{$E'$ does not connect $x$ to $y$}
         \end{scarray}
        }
   \!\!\!\!\!  q^{k(E')} \;  \prod_{e \in E'}  v_e
\end{eqnarray}
Note that $\zgxconny(\qve)$ and $\zgxnoconny(\qve)$ are polynomials in $q$
and $\{ v_e \}$:
the lowest-order contribution to $\zgxconny$ (resp.\ $\zgxnoconny$)
is of order at least $q$ (resp.\ at least $q^2$);
the highest-order contribution to $\zgxconny$
is of order at most $q^{|V|-{\rm dist}(x,y)}$,
where ${\rm dist}(x,y)$ is the length of the shortest path
from $x$ to $y$ using edges having $v_e \neq 0$
(if no such path exists, then $\zgxconny$ is identically zero);
and the highest-order contribution to $\zgxnoconny$
is of order exactly $q^{|V|}$
(with coefficient 1, coming from the term $E' = \emptyset$).
{}From \reff{eq1.2} we have trivially
\be
   Z_G(\qve)  \;=\;  \zgxconny(\qve) \,+\, \zgxnoconny(\qve)
 \label{eq2.G}
\ee
and almost as trivially
\be
   Z_{G \bullet xy}(\qve)  \;=\;  \zgxconny(\qve) \,+\, q^{-1} \zgxnoconny(\qve)
   \;.
 \label{eq2.Gxy}
\ee

\medskip
\noindent
{\bf Remark.}  Let $G + xy$ denote the graph $G$ with an extra edge $xy$ added.
Then it is not hard to see that \reff{eq1.2} also implies
\be
   Z_{G + xy}(\qve, v_{xy})  \;=\;  (1+v_{xy}) \zgxconny(\qve) \,+\,
      \left( 1 + {v_{xy} \over q} \right) \zgxnoconny(\qve)
   \;,
 \label{eq2.G+xy}
\ee
which equals $Z_G + v_{xy} Z_{G \bullet xy}$
in agreement with the deletion-contraction formula.
In particular, when $v_{xy} = -1$ we have
\be
   Z_{G + xy}(\qve, v_{xy} = -1)  \;=\;
      {q-1 \over q} \, \zgxnoconny(\qve)
   \;,
 \label{eq2.G+xy-1}
\ee
which equals $Z_G - Z_{G \bullet xy}$.

\bigskip
\smallskip

Let now $q$ be an integer $\ge 1$, and define the restricted partition function
\be
   Z_{G,x,y}(q, \{v_e\}; \sigma_x, \sigma_y)   \;=\;
   \sum_{ \begin{scarray}
              \sigma \colon\, V \to \{1,\ldots,q\} \\
              \sigma(x) = \sigma_x \\
              \sigma(y) = \sigma_y
          \end{scarray} }
      \,  \prod_{e \in E}  \,
      \biggl[ 1 + v_e \delta(\sigma(x_1(e)), \sigma(x_2(e))) \biggr]
 \label{Z_restricted}
\ee
where $\sigma_x, \sigma_y \in \{1,\ldots,q\}$
and the sum runs over all maps $\sigma\colon\, V \to \{1,\ldots,q\}$
satisfying $\sigma(x) = \sigma_x$ and $\sigma(y) = \sigma_y$.
We then have the following refinement of the
Fortuin-Kasteleyn identity \reff{eq1.2}:

\begin{proposition}
  \label{prop2.1}
\be
   Z_{G,x,y}(q, \{v_e\}; \sigma_x, \sigma_y)   \;=\;
   A_{G,x,y}(q, \{v_e\})  \,+\,
      B_{G,x,y}(q, \{v_e\}) \, \delta(\sigma_x,\sigma_y)
 \label{eq_prop2.1_1}
\ee
where
\begin{subeqnarray}
   A_{G,x,y}(q, \{v_e\})   & = &  q^{-2} \zgxnoconny(\qve)  \\[2mm]
   B_{G,x,y}(q, \{v_e\})   & = &  q^{-1} \zgxconny(\qve)
 \label{eq_prop2.1_2}
\end{subeqnarray}
are polynomials in $q$ and $\{v_e\}$, whose degrees in $q$ are
\begin{subeqnarray}
   \deg A   & = &  |V| \,-\, 2  \\
   \deg B   &\le&  |V| \,-\, 1 \,-\, {\rm dist}(x,y)
\end{subeqnarray}
where ${\rm dist}(x,y)$ is the length of the shortest path
from $x$ to $y$ using edges having $v_e \neq 0$
(if no such path exists, then $B = 0$).
Moreover, (\ref{eq_prop2.1_2}a/b) define the {\em unique}\/
functions $A_{G,x,y}$ and $B_{G,x,y}$ that are
polynomials in $q$ and satisfy \reff{eq_prop2.1_1}.
\end{proposition}

\par\medskip\noindent
{\sc First Proof.\ }
In \reff{Z_restricted}, expand out the product over $e \in E$,
and let $E' \subseteq E$ be the set of edges for which the term
$v_e \delta(\sigma(x_1(e)), \sigma(x_2(e)))$ is taken.
Now perform the sum over configurations
$\{ \sigma(z) \} _{z \in V \setminus \{x,y\}}$:
in each connected component of the subgraph $(V,E')$
the spin value $\sigma(z)$ must be constant.
In particular, in each component containing $x$ and/or $y$,
the spins must all equal the specified value $\sigma_x$ and/or $\sigma_y$;
in all other components, the spin value is free.
Therefore,
\be
   Z_{G,x,y}(q, \{v_e\}; \sigma_x, \sigma_y)   \;=\;
   \cases{ q^{-2} \zgxnoconny(\qve)   & if $\sigma_x \neq \sigma_y$  \cr
          \noalign{\vskip 2mm}
           q^{-2} \zgxnoconny(\qve) + q^{-1} \zgxconny(\qve)
                                      & if $\sigma_x = \sigma_y$  \cr
         }
\ee
Finally, the numbers $A_{G,x,y}$ and $B_{G,x,y}$ in \reff{eq_prop2.1_1}
are uniquely defined for each integer $q \ge 2$;
and any polynomials that coincide with (\ref{eq_prop2.1_2}a/b)
at all integers $q \ge 2$ must coincide everywhere.
\qed

\par\medskip\noindent
{\sc Second Proof.\ }
For each integer $q$, the $S_q$ permutation symmetry of the Potts model
implies that \be
   Z_{G,x,y}(q, \{v_e\}; \sigma_x, \sigma_y)   \;=\;
      A \,+\, B \delta(\sigma_x,\sigma_y)
 \label{eq.secondproof}
\ee
for some numbers $A$ and $B$ depending on $G$, $x$, $y$, $q$ and $\{ v_e \}$;
moreover, the numbers $A$ and $B$ are obviously unique when $q \ge 2$.
But then, summing \reff{eq.secondproof} over $\sigma_x, \sigma_y$
without and with the constraint $\sigma_x = \sigma_y$, we get
\begin{subeqnarray}
   Z_G                & = &  q^2 A \,+\, q B   \\
   Z_{G \bullet xy}   & = &  q A \,+\, q B
\end{subeqnarray}
Hence
\begin{subeqnarray}
   A   & = &   {Z_G - Z_{G \bullet xy}   \over   q(q-1)}
       \;=\;   q^{-2} \zgxnoconny
   \\[2mm]
   B   & = &   {q Z_{G \bullet xy} - Z_G   \over   q(q-1)}
       \;=\;   q^{-1} \zgxconny
 \label{eq2.AB}
\end{subeqnarray}
by virtue of \reff{eq2.G} and \reff{eq2.Gxy}.
\qed

\medskip
\noindent
{\bf Remark.}  Extensions of Proposition~\ref{prop2.1}
to $k$-rooted graphs (for any $k \ge 2$) can also be derived
\cite{Wu_99,Polin}.
I thank Fred Wu for bringing reference \cite{Wu_99} to my attention.

\bigskip
\smallskip

Let us now consider inserting the 2-rooted graph $(G,x,y)$
in place of an edge $e_*$ in some other graph $H$,
and let us call the resulting graph $H^G$.
We can trivially rewrite \reff{eq_prop2.1_1}/\reff{eq_prop2.1_2} as
\begin{subeqnarray}
   Z_{G,x,y}(q, \{v_e\}; \sigma_x, \sigma_y)   & = &
   A_{G,x,y}(q, \{v_e\})
   \left[ 1  \,+\,
          {B_{G,x,y}(q, \{v_e\}) \over A_{G,x,y}(q, \{v_e\})}
          \, \delta(\sigma_x,\sigma_y)
   \right]
      \\[2mm]
   & = &
   A_{G,x,y}(q, \{v_e\})
   \left[ 1  \,+\, v_{\hboxscript{eff,G,x,y}}(\qve)
          \, \delta(\sigma_x,\sigma_y)
   \right]
   \;,
\end{subeqnarray}
where
\be
   v_{\hboxscript{eff}}  \;\equiv\;
   v_{\hboxscript{eff,G,x,y}}(\qve)
    \;=\;  {B_{G,x,y}(\qve) \over A_{G,z,y}(\qve)}
    \;=\;  {q \zgxconny(\qve) \over \zgxnoconny(\qve)}
 \label{veff_def}
\ee
is a rational function of $q$ and $\{v_e\}$.
We then see that $(G,x,y)$ acts within the graph $H$
as a single edge with effective weight $v_{\hboxscript{eff}}$,
provided that the partition function $Z_H$ is multiplied by
an overall prefactor $A_{G,z,y}(\qve)$.
More precisely,
\be
   Z_{H^G}(q, \{v_e\}_{e \in (E(H) \setminus e_*) \cup E(G)})
   \;=\;
   A_{G,x,y}(q, \{v_e\}_{e \in E(G)})
   \,
   Z_H(q, \{v_e\}_{e \in E(H) \setminus e_*}, v_{\hboxscript{eff}})
   \;,
\ee
where $v_{\hboxscript{eff}} \equiv
       v_{\hboxscript{eff,G,x,y}}(q,\{v_e\}_{e \in E(G)})$
replaces $v_{e_*}$ as an argument of $Z_H$.
This follows from Proposition~\ref{prop2.1}
whenever $q$ is an integer $\ge 1$;
and the corresponding identity then holds for all $q$,
because both sides are rational functions of $q$
that agree at infinitely many points.
It is also worth noting that the ``transmissivity''
$t_{\hboxscript{eff}} \equiv v_{\hboxscript{eff}} / (q + v_{\hboxscript{eff}})$
is given by the simple formula
\be
   t_{\hboxscript{eff}}
%%    \;=\;  {B_{G,x,y}(\qve) \over q A_{G,x,y}(\qve) + B_{G,x,y}(\qve)}
    \;=\;  {\zgxconny(\qve) \over Z_G}
   \;.
\ee

The most general version of this construction appears to be the following
\cite{Woodall_00}:
Let $H=(V,E)$ be a finite undirected graph,
and let $\vec{H}$ be a directed graph obtained by assigning an orientation
to each edge of $H$.
For each edge $e \in E$, let $G_e = (V_e, E_e, x_e, y_e)$
be a 2-rooted finite undirected graph
(so that $x_e, y_e \in V_e$ with $x_e \neq y_e$)
equipped with edge weights $\{ v_{\widetilde{e}} \} _{\widetilde{e} \in E_e}$.
We denote by ${\bf G}$ the family $\{ G_e \} _{e \in E}$,
and we denote by $\vec{H}^{\bf G}$ the undirected graph obtained from $H$
by replacing each edge $e \in E$ with a copy of the corresponding graph $G_e$,
attaching $x_e$ to the tail of $e$ and $y_e$ to the head.
Its edge set is thus ${\bf E} = \bigcup\limits_{e \in E} E_e$ (disjoint union).
We then have:

\begin{proposition}
   \label{prop2.2}
Let $H=(V,E)$ and $\{ G_e \} _{e \in E}$ be as above.
Suppose that
\be
   Z_{G_e,x_e,y_e}(q, \{v_{\widetilde{e}} \} _{\widetilde{e} \in E_e};
                   \sigma_{x_e}, \sigma_{y_e})
   \;=\;
   A_{G_e,x_e,y_e}(q, \{v_{\widetilde{e}} \} _{\widetilde{e} \in E_e})
   \,+\,
   B_{G_e,x_e,y_e}(q, \{v_{\widetilde{e}} \} _{\widetilde{e} \in E_e}) \,
      \delta(\sigma_{x_e},\sigma_{y_e})
   \;,
\ee
and define
\be
   v_{\hboxscript{eff},e}  \;\equiv\;
    {B_{G_e,x_e,y_e}(q, \{v_{\widetilde{e}} \} _{\widetilde{e} \in E_e})
     \over
     A_{G_e,x_e,y_e}(q, \{v_{\widetilde{e}} \} _{\widetilde{e} \in E_e})
    }
   \;.
\ee
Then
\be
   Z_{\vec{H}^{\bf G}}(q, \{v_{\widetilde{e}} \} _{\widetilde{e} \in {\bf E}})
   \;=\;
   \left( \prod\limits_{e \in E}
          A_{G_e,x_e,y_e}(q, \{v_{\widetilde{e}} \} _{\widetilde{e} \in E_e})
   \right)
   \,\times\,
   Z_H(q, \{v_{\hboxscript{eff},e} \} _{e \in E} )
   \;.
\ee
In particular, $Z_{\vec{H}^{\bf G}}$ does not depend on the orientations
of the edges of $H$.
\end{proposition}

\noindent
A special case of this construction is the subdivision of edges
\cite{Read_99,Read_00,Read_03,Brown-Hickman_99b},
discussed further in Appendix~\ref{app.A}.

\subsection{Parallel and series connection}   \label{sec2.2}

Important special cases of Proposition~\ref{prop2.1}
concern parallel and series connection.

The case of parallel connection is almost trivial:
Let $G$ consist of edges $e_1,\ldots,e_n$
(with corresponding weights $v_1,\ldots,v_n$)
in parallel between the same pair of vertices $x,y$.
Then
\be
   \prod\limits_{i=1}^n \left[1 + v_i \delta(\sigma_x,\sigma_y) \right]
   \;=\;
   1 \,+\, \left[ \prod\limits_{i=1}^n (1+v_i) \,-\, 1 \right]
   \delta(\sigma_x,\sigma_y)
   \;,
 \label{eq2.parallel}
\ee
so that the parallel edges are equivalent to a single edge
with effective weight
\be
   \veff  \;=\;  \prod\limits_{i=1}^n (1+v_i) \,-\, 1
 \label{eq2.parallel2}
\ee
or equivalently
\be
   1 + \veff  \;=\;  \prod\limits_{i=1}^n (1+v_i)
   \;.
\ee
That is, for parallel connection, ``$1+v$ multiplies''.

The case of series connection is a bit less trivial:
Let $G$ be a path $P_n$ consisting of edges $e_1,\ldots,e_n$
(with weights $v_1,\ldots,v_n$) between endvertices 0 and $n$.
Then
\be
   \sum\limits_{\sigma_1,\ldots,\sigma_{n-1} = 1}^q
   \prod\limits_{i=1}^n \left[1 + v_i \delta(\sigma_{i-1},\sigma_i) \right]
   \;=\;
   A \,+\, B \delta(\sigma_0,\sigma_n)
 \label{eq2.series}
\ee
with
\begin{subeqnarray}
   A & = &  q^{-1} \left[ \prod\limits_{i=1}^n (q+v_i)
                          \,-\, \prod\limits_{i=1}^n v_i  \right]
   \slabel{eq2.seriesABa}   \\[2mm]
   B & = &  \prod\limits_{i=1}^n v_i
 \label{eq2.seriesAB}
\end{subeqnarray}
This formula can easily be proven by induction on $n$;
or it can alternatively be derived from \reff{eq2.AB} by remembering that
\begin{eqnarray}
   Z_G \;=\; Z_{P_n}  & = &  q \prod\limits_{i=1}^n (q+v_i)   \\[2mm]
   Z_{G \bullet 0n} \;=\; Z_{C_n}  & = &
     \prod\limits_{i=1}^n (q+v_i) \,+\, (q-1) \prod\limits_{i=1}^n v_i
    \label{eq2.ZCn}
\end{eqnarray}
(formulae that can themselves be proven by induction).
In particular, the series edges are equivalent to a single edge
with effective weight
\be
   \veff \;\equiv\;  {B \over A}  \;=\;
      {q \prod\limits_{i=1}^n v_i  \over
       \prod\limits_{i=1}^n (q+v_i) \,-\, \prod\limits_{i=1}^n v_i }
 \label{veff_series_1}
\ee
or equivalently
\be
   {\veff \over q+\veff}   \;=\;  \prod\limits_{i=1}^n {v_i \over q+v_i}
   \;.
 \label{veff_series_2}
\ee
That is, for series connection, ``$v/(q+v)$ multiplies''.
But one should not forget the overall prefactor $A$
given by \reff{eq2.seriesABa}.

\bigskip

\noindent
{\bf Remarks.}
1.  Note that applying series reduction
\reff{veff_series_1}/\reff{veff_series_2}
to a chromatic polynomial (all $v_i = -1$) leads to edge weights $\veff$
that are in general unequal and $\neq -1$.
This is further evidence of the value of studying the
general Potts-model partition function $Z_G(\qve)$
with not-necessarily-equal edge weights $\{v_e\}$.
In particular, the series-reduction formulae
allow an easy proof of the Brown--Hickman theorem on chromatic roots of
large subdivisions \cite{Brown-Hickman_99b}: see Appendix~\ref{app.A}.

2.  The foregoing formulae form the basis for a linear-time
[i.e.\ $O(|V| + |E|)$] algorithm for computing
the Potts-model partition function $Z_G(\qve)$
for a series-parallel graph $G$
at any fixed set of numbers $q$ and $\{v_e\}$,
and hence
%%  just compute it at $q=0,1,\ldots,n-1$ and interpolate
a quadratic-time [$O(|V|^2 + |V| |E|)$] algorithm
for computing $Z_G(\qve)$ as a polynomial in $q$ for fixed $\{v_e\}$.\footnote{
   This is in a computational model where the elementary arithmetic
   operations ($+$, $-$, $\times$, $\div$) are assumed to take
   a time of order 1 irrespective of the size of their arguments.
}
(One needs, as a subroutine, a linear-time algorithm
 for recognizing whether a graph is series-parallel and,
 if it is, computing a decomposition tree of series-parallel reductions:
 see \cite[Sections 2.3 and 3.3]{Valdes_82}.)
This is essentially the approach used, for the special case of
the chromatic polynomial, in \cite{Chandrasekharan_94}.
%% See also Oxley--Welsh \cite{Oxley_92}.
More generally, Noble \cite{Noble_98} has shown how to compute,
in linear time, the dichromatic polynomial $Z_G(q,v)$
at any point $(q,v)$ for graphs of bounded tree-width
(the case of series-parallel graphs corresponds to tree-width 2);
this algorithm presumably extends to handle the
general Potts-model partition function at any point $(q,\{v_e\})$.
%%
%% See letters.to-from.noble.May_4_00
%%
For general graphs, by contrast ---
or even for general {\em planar}\/ graphs of maximum degree 4 ---
even the problem of determining whether $G$ is 3-colorable
[i.e.\ whether $P_G(3) \neq 0$] is NP-complete \cite[p.~191]{Garey_79}.

\subsection{Potts-model partition function of generalized theta graphs}
   \label{sec2.3}

Consider a generalized theta graph $\Theta_{s_1,\ldots,s_p}$
consisting of endvertices $x,y$ connected by $p$ internally disjoint paths
of lengths $s_1, \ldots, s_p \ge 1$.
Label the edges $e_{ij}$ ($1 \le i \le p$, $1 \le j \le s_i$)
and let $v_{ij}$ be the corresponding weights.
Applying the series-connection identity \reff{eq2.series}/\reff{eq2.seriesAB}
on each of the $p$ paths,
and then applying the parallel-connection identity \reff{eq2.parallel},
we obtain the restricted partition function
\be
   Z_{\Theta_{s_1,\ldots,s_p},x,y}(q, \{v_{ij}\}; \sigma_x,\sigma_y)
   \;=\;
   A \,+\, B \delta(\sigma_x,\sigma_y)
 \label{eq2.theta}
\ee
with
\begin{subeqnarray}
   A  & = &  q^{-p}  \prod\limits_{i=1}^p
    \left[ \prod\limits_{j=1}^{s_i} (q + v_{ij})
           \,-\,  \prod\limits_{j=1}^{s_i} v_{ij}  \right]       \\[4mm]
   B  & = &  q^{-p}  \left\{
    \prod\limits_{i=1}^p
       \left[ \prod\limits_{j=1}^{s_i} (q + v_{ij})
              \,+\, (q-1)  \prod\limits_{j=1}^{s_i} v_{ij}  \right]
       \;-\;
    \prod\limits_{i=1}^p
       \left[ \prod\limits_{j=1}^{s_i} (q + v_{ij})
              \,-\,  \prod\limits_{j=1}^{s_i} v_{ij}  \right]
    \right\}
    \nonumber \\
 \label{eq2.thetaAB}
\end{subeqnarray}
In particular, summing \reff{eq2.theta} over $\sigma_x$ and $\sigma_y$
without constraint, we obtain $Z_{\Theta_{s_1,\ldots,s_p}} = q^2 A + qB$:

\begin{proposition}
   \label{prop2.3}
The Potts-model partition function for the generalized theta graph
$\Theta_{s_1,\ldots,s_p}$ with edge weights
$\{ v_{ij} \} _{1 \le i \le p, \, 1 \le j \le s_i}$ is
\begin{eqnarray}
    & &  \!\!\!\!
         Z_{\Theta_{s_1,\ldots,s_p}}(q, \{v_{ij}\})   \nonumber \\[2mm]
    & & \quad=\;  q^{-(p-1)} \left\{
    \prod\limits_{i=1}^p
       \left[ \prod\limits_{j=1}^{s_i} (q + v_{ij})
              + (q-1)  \prod\limits_{j=1}^{s_i} v_{ij}  \right]
       \,+\, (q-1)
    \prod\limits_{i=1}^p
       \left[ \prod\limits_{j=1}^{s_i} (q + v_{ij})
              -  \prod\limits_{j=1}^{s_i} v_{ij}  \right]
    \right\}
    \;.
    \nonumber \\
 \label{eq2.thetaZ}
\end{eqnarray}
In particular, when $v_{ij} = v$ for all $i,j$, we have the dichromatic
polynomial
\be
    Z_{\Theta_{s_1,\ldots,s_p}}(q,v)   \;=\;
    q^{-(p-1)} \left\{
    \prod\limits_{i=1}^p
       \left[ (q+v)^{s_i} + (q-1) v^{s_i} \right]
       \,+\, (q-1)
    \prod\limits_{i=1}^p
       \left[ (q+v)^{s_i} - v^{s_i} \right]
    \right\}
    \;.
 \label{eq2.thetaZqv}
\ee
And when also $s_1 = \ldots = s_p = s$, we have
\be
   Z_{\Theta^{(s,p)}}(q,v)   \;=\;
   q^{-(p-1)} \left\{
       \left[ (q+v)^s + (q-1) v^s \right]^p
       \,+\, (q-1)
       \left[ (q+v)^s - v^s \right]^p
    \right\}
    \;.
 \label{eq2.thetaZqvs}
\ee
\end{proposition}

\medskip
\noindent
{\bf Remarks.}
1.  Here is an alternate derivation of
\reff{eq2.theta}/\reff{eq2.thetaAB} and \reff{eq2.thetaZ}:
If in $\Theta_{s_1,\ldots,s_p}$ we contract $x$ to $y$,
we obtain $p$ cycles of lengths $s_i$ joined at a single vertex, so that
\begin{eqnarray}
   Z_{\Theta_{s_1,\ldots,s_p} \bullet xy}
      (q, \{ v_{ij} \} _{1 \le i \le p, \, 1 \le j \le s_i})
   & = &
   q^{-(p-1)} \prod\limits_{i=1}^p
                  Z_{C_{s_i}}(q, \{ v_{ij} \} _{1 \le j \le s_i})
   \nonumber \\[2mm]
   & = &  q^{-(p-1)} \prod\limits_{i=1}^p
       \left[ \prod\limits_{j=1}^{s_i} (q + v_{ij})
              \,+\, (q-1)  \prod\limits_{j=1}^{s_i} v_{ij}  \right]
   \;.
   \nonumber \\
\end{eqnarray}
If, on the other hand, we add to $\Theta_{s_1,\ldots,s_p}$
an extra edge $xy$ with $v_{xy} = -1$, we obtain $p$ cycles
of lengths $s_i +1$ joined along a single $v=-1$ edge, so that
\begin{eqnarray}
   & &  \!\!\!\!
      Z_{\Theta_{s_1,\ldots,s_p} + xy}
      (q, \{ v_{ij} \} _{1 \le i \le p, \, 1 \le j \le s_i}, v_{xy}=-1)
   \nonumber \\[2mm]
   & & \qquad\qquad=\;
      [q(q-1)]^{-(p-1)} \prod\limits_{i=1}^p
                  Z_{C_{s_i+1}}(q, \{ v_{ij} \} _{1 \le j \le s_i}, v_{xy}=-1)
   \nonumber \\[2mm]
   & & \qquad\qquad=\;
     [q(q-1)]^{-(p-1)} \prod\limits_{i=1}^p
       \left[ (q-1) \prod\limits_{j=1}^{s_i} (q + v_{ij})
              \,-\, (q-1)  \prod\limits_{j=1}^{s_i} v_{ij}  \right]
   \nonumber \\[2mm]
   & & \qquad\qquad=\;
     q^{-(p-1)} (q-1) \prod\limits_{i=1}^p
       \left[ \prod\limits_{j=1}^{s_i} (q + v_{ij})
              \,-\, \prod\limits_{j=1}^{s_i} v_{ij}  \right]
   \;.
 \label{eq2.thetaZ+xy}
\end{eqnarray}
The addition-contraction formula
$Z_G = Z_{G \bullet xy} + Z_{G+xy}(v_{xy}=-1)$
then gives \reff{eq2.thetaZ};
and using \reff{eq2.AB} we obtain \reff{eq2.thetaAB}.

2. In Appendix~\ref{app.B} I use formula \reff{eq2.thetaAB}
to give an extension (and simple proof)
of Thomassen's \cite[Theorem 3.9]{Thomassen_97}
construction concerning the chromatic roots of 2-degenerate graphs.

\section{Limit Sets of Zeros for Certain Sequences of Analytic Functions}
   \label{sec3}

Let $(f_n)$ be a sequence of analytic functions on a domain $D \subset \C$,
and let $(a_n)$ be a sequence of positive real numbers
such that $(|f_n|^{a_n})$ are uniformly bounded on compact subsets of $D$.
We are interested in the zero sets $\scrz(f_n)$
and in particular in their limit sets as $n\to\infty$.
We shall relate these sets to the existence and behavior of the limit
\be
   u(z)  \;=\;  \lim_{n\to\infty}  a_n \log |f_n(z)|  \;.
 \label{eq3.1}
\ee
We begin with a warm-up result:

\begin{theorem}
   \label{thm3.1}
Let $D$ be a domain in $\C$, and let $x_0 \in D \cap \R$.
Let $(f_n)$ be analytic functions on $D$,
and let $(a_n)$ be positive real constants
such that $(|f_n|^{a_n})$ are uniformly bounded on compact subsets of $D$.
Assume further that
\begin{itemize}
   \item[(a)]  For each $x \in D \cap \R$, $f_n(x)$ is real and $> 0$.
   \item[(b)]  For each $x \in D \cap \R$,
     $\lim\limits_{n\to\infty}   a_n \log f_n(x) \equiv u(x)$
     exists and is finite.
   \item[(c)]  $u$ is {\em not}\/ real-analytic at $x_0$.
\end{itemize}
Then $\liminf \scrz(f_n) \ni x_0$:
that is, for all $n$ sufficiently large,
there exist zeros $z^*_n$ of $f_n$
such that $\lim\limits_{n\to\infty} z^*_n = x_0$.
\end{theorem}

Theorem \ref{thm3.1} is well-known to workers in mathematical statistical
mechanics:  it is the contrapositive of an observation going back to
Yang and Lee \cite{Yang-Lee_52} concerning the genesis of phase transitions
(see e.g.\ \cite[Theorem 4.1, p.~51]{Griffiths_72}).
%%  {\bf See also Ruelle, Stat Mech, Exercise 3.G, page 69.}
We give its proof later in this section.

%%  \bigskip
%%  
%%  {\bf HERE IS AN ALTERNATE VERSION OF THEOREM \ref{thm3.1}:
%%    Suppose that there exists a connected set $A \subset D$
%%    and continuous functions $g_n \colon\, A \to \C$
%%    such that $f_n = \exp(g_n)$ on $A$.
%%    [In particular, it is necessary that the $f_n$ be nonvanishing on $A$;
%%     and if $A$ is simply connected [what does this mean for a general set??],
%%     then it's sufficient.]
%%    Let $\Sfin \subset A$ be the set where the limit
%%    $$ g(z) = \lim_{n\to\infty} a_n g_n(z) $$
%%    exists and is finite.
%%    NOW WHAT????}
%%  
%%  
%%  \bigskip

The main result of this section, Theorem \ref{thm3.2},
is similar in nature to Theorem \ref{thm3.1}:
the difference is that, by taking the logarithm only of the
{\em absolute value}\/ of $f_n(z)$,
we need not worry {\em a priori}\/ whether $f_n(z) \neq 0$
(since $\log 0 = -\infty$ is a legitimate value),
nor need we worry about potential ambiguities in the branch of the logarithm.
%%  As compensation, we have to demand convergence of \reff{eq3.1}
%%  on a somewhat larger subset of the domain $D$.
%%  {\bf Is this really true anymore???
%%     Do we need to introduce the notion of copiousness at all???}
%%  Let us say that a subset $S \subset D$ is {\em copious at $z \in D$}\/
%%  in case, for all open neighborhoods $U \ni z$,
%%  every harmonic function on $U$ that vanishes on $U \cap S$
%%  is identically vanishing.
%%  {\bf Is this concept known in harmonic-function theory???}

\begin{theorem}
   \label{thm3.2}
Let $D$ be a domain in $\C$, and let $z_0 \in D$.
Let $(f_n)$ be analytic functions on $D$,
and let $(a_n)$ be positive real constants
such that $(|f_n|^{a_n})$ are uniformly bounded on compact subsets of $D$.
% Define
% \be
%    \Sfin   \;=\;
%    \{ z \in D \colon\;  u(z) \equiv \lim_{n\to\infty}  a_n \log |f_n(z)|
%                    \hbox{ exists and is finite} \}
%    \;.
% \ee
Suppose that there does {\em not}\/ exist a neighborhood $U \ni z_0$
and a function $v$ on $U$ that is either harmonic or else identically $-\infty$
such that $\liminf\limits_{n\to\infty} a_n \log |f_n(z)| \le v(z)
 \le \limsup\limits_{n\to\infty} a_n \log |f_n(z)|$ for all $z \in U$.
Then $z_0 \in \liminf \scrz(f_n)$:
that is, for all $n$ sufficiently large,
there exist zeros $z^*_n$ of $f_n$
such that $\lim\limits_{n\to\infty} z^*_n = z_0$.
\end{theorem}

\noindent

The proofs of Theorems~\ref{thm3.1} and \ref{thm3.2}
depend crucially on the
concept of a {\em normal family}\/ of analytic functions
\cite{Montel_27,Schiff_93}:
\begin{definition}
   \label{def_normal}
A family $\scrf$ of analytic functions on a domain $D \subset \C$
is said to be {\bf normal} (in $D$) if, for every sequence
$(f_n) \subset \scrf$, there exists a subsequence $(f_{n_k})$ that converges,
uniformly on compact subsets of $D$, either to a (finite-valued)
analytic function or else to the constant function $\infty$.
\end{definition}
(Some authors omit the possibility of convergence to $\infty$
from the definition of normality.  But the foregoing definition
is more useful for our purposes.)
An easy covering and diagonalization argument \cite[Theorem 2.1.2]{Schiff_93}
shows that normality is a local property:
$\scrf$ is normal in $D$ if and only if, for each $z \in D$,
there exists an open disc $U \ni z$ in which $\scrf$ is normal.

The first key result concerning normal families is Montel's (1907) theorem
\cite[pp.~35--36]{Schiff_93}:  if the family $\scrf$ is uniformly bounded
on compact subsets of $D$, then it is normal.
(In this case, of course, convergence to $\infty$ is impossible.)
We will need a slight extension of this result
\cite[Example 2.3.9]{Schiff_93}.
For each set $A \subset \C$,
let $\scrf_{\notin A}$ be the family of analytic functions
whose values avoid $A$:
\be
   \scrf_{\notin A} \;=\; \{f \hbox{ analytic on } D \colon\;
                            f[D] \cap A = \emptyset \}
   \;.
\ee
We then have:
\begin{proposition}
   \label{prop_normality}
If $A \subset \C$ has nonempty interior,
then $\scrf_{\notin A}$ is normal.
\end{proposition}

{\bf Remarks.}
1.  Since normality is a local property,
this result can be strengthened as follows:
for a family $\scrf$ to be normal,
it suffices that for each $z \in D$
there exist an open disc $U \ni z$
and a set $A_U \subset \C$ with nonempty interior,
such that $f[U] \cap A_U = \emptyset$ for all $f \in \scrf$.

2.  A result vastly stronger than Proposition~\ref{prop_normality} is true:
Montel's {\em Crit\`ere fondamental}\/ (1912) states that
$\scrf_{\notin A}$ is normal as soon as $A$ contains
{\em at least two points}\/.
Detailed accounts of this deep result can be found
in \cite{Montel_27,Schiff_93}.
But we won't need it.

\bigskip
\noindent
{\sc Proof of Proposition~\ref{prop_normality}.\ }
Suppose that $A$ contains the disc $\{ w \colon\; |w-a| < \epsilon\}$.
For each $f \in \scrf_{\notin A}$, define $\widetilde{f}(z) = 1/[f(z)-a]$.
Then the family
$\widetilde{\scrf} \equiv
 \{ \widetilde{f} \colon\;  f \in \scrf_{\notin A} \}$
is uniformly bounded by $1/\epsilon$, hence normal by Montel's theorem.
So let $(f_n) \subset \scrf_{\notin A}$,
and consider the corresponding sequence
$(\widetilde{f}_n) \subset \widetilde{\scrf}$.
By normality of $\widetilde{\scrf}$,
there exists a subsequence $(\widetilde{f}_{n_k})$ that converges,
uniformly on compact subsets of $D$, to an analytic function $g$.
But since the $\widetilde{f}_{n_k}$ are nonvanishing on $D$,
Hurwitz's theorem \cite[p.~262]{Remmert_91} tells us that
$g$ is either identically zero or else nonvanishing.
It is straightforward to show that,
in the former case,
the corresponding subsequence $(f_{n_k})$ converges,
uniformly on compact subsets of $D$,
to the constant function $\infty$;
and that in the latter case,
$(f_{n_k})$ converges,
uniformly on compact subsets of $D$,
to the analytic function $f = a + 1/g$.
(For each compact $K \subset D$, we have
 $\inf\limits_{z \in K} |g(z)| \equiv \delta > 0$,
so for all sufficiently large $k$ we have
$|\widetilde{f}_{n_k}| \ge \delta/2$ everywhere on $K$.
Now apply the uniform continuity of the function $w \mapsto 1/w$
on the set where $|w| \ge \delta/2$.)
\qed

We also need the following well-known lemma,
which expresses the main idea underlying the Vitali-Porter theorem:

\begin{lemma}
   \label{lemma_Vitali}
Let $\scrf$ be a normal family of analytic functions
on a domain $D \subset \C$,
and suppose that the sequence $(f_n) \subset \scrf$
converges (pointwise either to a complex number or to $\infty$)
on a set $S \subset D$ having at least one accumulation point in $D$.
Then $(f_n)$ converges, uniformly on compact subsets of $D$,
either to a (finite-valued) analytic function
or else to the constant function $\infty$.
\end{lemma}

\proof
Define $f(z) = \lim_{n \to\infty} f_n(z)$ for $z \in S$.
Now, for any subsequence $\scrs = (f_{n_k})$,
there exists (by normality) a subsubsequence $\scrs' = (f_{n_{k_l}})$
that converges, uniformly on compact subsets of $D$,
either to a (finite-valued) analytic function $g_{\scrs'}$
or else to the constant function $g_{\scrs'} \equiv \infty$.
Moreover, we must have $g_{\scrs'} \restrict S = f$ for all $\scrs'$.
Therefore, we either have $f \equiv \infty$, in which case
$g_{\scrs'} \equiv \infty$ for all $\scrs'$;
or else $f$ is everywhere finite-valued, in which case the $g_{\scrs'}$
are all equal to the {\em same}\/ analytic function $g$
(since $S$ is a determining set for analytic functions on $D$).
It follows that $(f_n)$ converges, uniformly on compact subsets of $D$,
either to the constant function $\infty$ (in the first case)
or to the analytic function $g$ (in the second).
\qed

We are now ready to prove Theorem~\ref{thm3.1}:

\bigskip\par\noindent
{\sc Proof of Theorem \ref{thm3.1}.\ }
Suppose the contrary, i.e.\ suppose that there exists an $\epsilon > 0$
and an infinite sequence $n_1 < n_2 < \ldots$ such that
none of the functions $f_{n_k}$ has a zero in the set
$D_\epsilon = \{ z \in \C \colon\; |z-x_0| < \epsilon \}$;
let us also take $\epsilon$ small enough so that
$\overline{D_\epsilon} \subset D$.
Then, since $D_\epsilon$ is simply connected,
$\log f_{n_k}$ is analytic in $D_\epsilon$
(we take the branch that is real on $D_\epsilon \cap \R$),
and $u_k \equiv a_{n_k} \log f_{n_k}$ satisfies
$\lim_{k\to\infty} u_k(x) = u(x)$ for $x \in D_\epsilon \cap \R$.
Moreover, by hypothesis the functions $\real u_k$ are
uniformly bounded above on $D_\epsilon$,
so it follows from Proposition~\ref{prop_normality}
that the $(u_k)$ are a normal family on $D_\epsilon$.
Lemma~\ref{lemma_Vitali} then implies that the sequence $(u_k)$ converges
on $D_\epsilon$ to an analytic function $\widetilde{u}$ that extends $u$.
But this contradicts the hypothesis that $u$ is not real-analytic at $x_0$.
\qed

Let us next recall that a real-valued function $u$
on a domain $D \subset \C \simeq \R^2$
is called {\em harmonic}\/ if it is twice continuously differentiable
and satisfies Laplace's equation
\be
   \Delta u \;\equiv\;
   {\partial^2 u \over \partial x^2} + {\partial^2 u \over \partial y^2}
   \;=\; 0   \;,
\ee
where we write $z = x+iy$.
The key facts (see e.g.\ \cite[pp.~122--124]{Cartan_63}) are:
the real part of every analytic function is harmonic;
and conversely, if the domain $D$ is simply connected,
then every harmonic function on $D$
is the real part of some analytic function.

One can define normality for families of harmonic functions
\cite[Section 5.4]{Schiff_93}:
\begin{definition}
   \label{def_normal_2}
A family $\scrh$ of harmonic functions on a domain $D \subset \C$
is said to be {\bf normal} (in $D$) if, for every sequence
$(u_n) \subset \scrh$, there exists a subsequence $(u_{n_k})$ that converges,
uniformly on compact subsets of $D$, either to a (finite-valued)
harmonic function or else to the constant function $+\infty$ or $-\infty$.
\end{definition}
A covering and diagonalization argument
analogous to \cite[Theorem 2.1.2]{Schiff_93}
shows that normality is a local property:
$\scrh$ is normal in $D$ if and only if, for each $z \in D$,
there exists an open disc $U \ni z$ in which $\scrh$ is normal.

There is a sufficient condition for normality analogous to
Proposition \ref{prop_normality}:
\begin{proposition}
   \label{prop_normality_2}
Let $\scrh$ be a family of harmonic functions on a domain $D$,
which are uniformly bounded above on compact subsets of $D$.
Then $\scrh$ is normal:
that is, for every sequence
$(u_n) \subset \scrh$, there exists a subsequence $(u_{n_k})$ that converges,
uniformly on compact subsets of $D$, either to a (finite-valued)
harmonic function or else to the constant function $-\infty$.
\end{proposition}

\proof
Since normality is a local property, it suffices to prove the
proposition when $D$ is an open disc $U$.
(The only property of $U$ we will really use is its simple connectedness.)
So let $(u_n)$ be a sequence of harmonic functions in $U$
that is uniformly bounded above on compact subsets of $U$.
Choose analytic functions $f_n$ on $U$ such that
$\real f_n = u_n$, and let $F_n = \exp(f_n)$.
The functions $F_n$ are nonvanishing,
and they are uniformly bounded on compact subsets of $U$.
By Montel's (1907) theorem,
there exists a subsequence $(F_{n_k})$ that converges,
uniformly on compact subsets of $U$,
to an analytic function $F$.
By Hurwitz's theorem, $F$ is either identically zero or else nonvanishing.
In the former case,
$u_{n_k} \equiv \real\log F_{n_k} = \log |F_{n_k}|$ tends to $-\infty$
uniformly on compact subsets of $U$.
In the latter case, $u_{n_k}$ tends to the harmonic function
$u_\infty \equiv \real \log F = \log |F|$
uniformly on compact subsets of $U$.
(For each compact $K \subset U$, we have
 $\inf\limits_{z \in K} |F(z)| \equiv \delta > 0$ and
 $\sup\limits_{z \in K} |F(z)| \equiv M < \infty$,
so for all sufficiently large $k$ we have
$\delta/2 \le |F_{n_k}| \le 2M$ everywhere on $K$.
Now apply the uniform continuity of the log function on $[\delta/2, 2M]$.)
\qed

\noindent
{\bf Remarks.}
1. For a slightly different proof,
see \cite[Theorems 5.4.2 and 5.4.3]{Schiff_93}.

2. Montel's proof \cite[Section 23]{Montel_27} is {\em not}\/ valid:
if a subsequence of $(f_n \equiv u_n + iv_n)$ tends to $\infty$,
it does not follow that the corresponding subsequence of $(|u_n|)$
also tends to $\infty$;  it could be that $(|v_n|)$ does so instead
(see e.g.\ \cite[p.~184]{Schiff_93}).

% \bigskip
% 
% We also need the harmonic-function analogue of Lemma~\ref{lemma_Vitali}:
% 
% \begin{lemma}
%    \label{lemma_Vitali_harmonic}
% Let $\scrh$ be a normal family of harmonic functions
% on a domain $D \subset \C$,
% and suppose that the sequence $(u_n) \subset \scrh$
% converges (pointwise either to a real number or to $+\infty$ or $-\infty$)
% on a set $S \subset D$ that is copious at at least one point $z_0 \in D$.
% Then $(u_n)$ converges, uniformly on compact subsets of $D$,
% either to a (finite-valued) harmonic function
% or else to the constant function $+\infty$ or $-\infty$.
% \end{lemma}
% 
% \proof
% Completely analogous to that of Lemma~\ref{lemma_Vitali}.
% \qed

\bigskip
\noindent
{\sc Proof of Theorem~\ref{thm3.2}.\ }
Suppose the contrary, i.e.\ suppose that there exists an $\epsilon > 0$
and an infinite sequence $n_1 < n_2 < \ldots$ such that
none of the functions $f_{n_k}$ has a zero in the set
$D_\epsilon = \{ z \in \C \colon\; |z-z_0| < \epsilon \} \subset D$.
Then each function $u_k \equiv a_{n_k} \log |f_{n_k}|$
is harmonic on $D_\epsilon$;
by hypothesis, the functions $u_k$ are
uniformly bounded above on compact subsets of $D_\epsilon$;
so it follows from Proposition~\ref{prop_normality_2}
that the $(u_k)$ are a normal family on $D_\epsilon$.
Therefore, there exists a subsequence $(u_{k_l})$ that converges,
uniformly on compact subsets of $D_\epsilon$,
to a function $v$ that is either harmonic on $D_\epsilon$
or else identically $-\infty$.
But this contradicts the hypothesis of the theorem.
%%  Moreover, we have by hypothesis that $\lim_{k\to\infty} u_k(z)$
%%  exists and is finite for all $z \in \Sfin \cap D_\epsilon$,
%%  a set which is copious at $z_0$.
%%  Lemma~\ref{lemma_Vitali_harmonic} then implies that
%%  the sequence $(u_k)$ converges on $D_\epsilon$
%%  to a (finite-valued) harmonic function $\widetilde{u}$ that extends $u$.
%%  But this contradicts the hypothesis that there does not exist
%%  a harmonic function $v$ on $D_\epsilon$ satisfying
%%  $\liminf\limits_{n\to\infty} a_n \log |f_n| \le v
%%  \le \limsup\limits_{n\to\infty} a_n \log |f_n|$.
\qed

\medskip
\noindent
{\bf Remark.}  An argument closely related to that of
Theorems~\ref{thm3.1} and \ref{thm3.2} was used previously,
in a special context,
by Borwein, Chen and Dilcher \cite[pp.~79 and 82]{Borwein_95}.
I~thank Karl Dilcher for informing me of this article.

\bigskip
\smallskip

We now proceed to the proof of Theorem~\ref{thm1.5}.
We have already defined
\be
   D_k  \;=\;  \{ z \colon\;  k \hbox{ is dominant at } z  \}
   \;.
\ee
Let us further define the ``good sets''
\be
   V_k  \;=\;  \{ z \colon\;  k \hbox{ is the unique dominant index at $z$,
                                     and $\alpha_k(z) \neq 0$} \}
 \label{def_Vk}
\ee
and the ``bad sets''
\begin{eqnarray}
   W_k  &=&  \{ z \colon\;  k \hbox{ is the unique dominant index at $z$,
                                     and $\alpha_k(z) =0$} \}     \\[4mm]
   X    &=&  \{ z \colon\;  \beta_1(z) = \beta_2(z) = \ldots = \beta_m(z) = 0 \}
                                                                  \\[4mm]
   Y_{kl}  &=&  \{ z \colon\;  \hbox{$k$ and $l$ ($k \neq l$) are dominant
                  indices at $z$ (there may be others)}        \nonumber \\
           & & \hspace{8.5cm} \hbox{and $|\beta_k(z)| > 0$} \}
   \label{def_Ykl}
\end{eqnarray}
Clearly all the sets \reff{def_Vk}--\reff{def_Ykl} are disjoint,
except that some of the $Y_{kl}$ may overlap at points where
there are three or more dominant indices.
We shall write $V = \bigcup\limits_{k=1}^m V_k$,
$W = \bigcup\limits_{k=1}^m W_k$ and
$Y = \!\bigcup\limits_{1 \le k < l \le m} Y_{kl}$;
we obviously have the disjoint decomposition
$D = V \cup W \cup X \cup Y$.
Theorem \ref{thm1.5} amounts to the statement that
\be
  \liminf \scrz(f_n)  \;=\; \limsup \scrz(f_n)  \;=\;  W \cup X \cup Y   \;.
\ee

We begin by collecting some simple facts about these sets:

\begin{lemma}
   \label{lemma3.8}
Under the hypotheses of Theorem~\ref{thm1.5}:
\begin{itemize}
   \item[(a)] Each $D_k$ is closed in $D$.  In particular, each $\partial D_k$
      has empty interior.
   \item[(b)] $D_k = V_k \cup W_k \cup X \cup
                 \Biggl( \bigcup\limits_{l \colon\, l \neq k} Y_{kl} \Biggr)$,
      and the latter four sets are disjoint.
   \item[(c)] For $k \neq l$, $D_k \cap D_l = X \cup Y_{kl}$.
      In particular, each set $X \cup Y_{kl}$ is closed.
   \item[(d)] For $k \neq l$, $Y_{kl} \subset \partial D_k \cap \partial D_l$.
      In particular, each $Y_{kl}$ has empty interior.
   \item[(e)] Each $V_k$ is open.
   \item[(f)] Each $V_k \cup W_k$ is open.  Moreover,
      $V_k \cup W_k \subset D_k^\circ \subset V_k \cup W_k \cup X$,
      where ${}^\circ$ denotes interior.
   \item[(g)] $W$ has no limit points in $D$.
      Moreover, for each $z \in W_k$ there exists $\epsilon > 0$ such that
      the punctured neighborhood $\{z' \colon\; 0 < |z'-z| < \epsilon \}$
      is contained in $V_k$.
   \item[(h)] $X$ has no limit points in $D$.
   \item[(i)] $W \cup X \cup Y$ is a closed set with empty interior,
      so that $V$ is a dense open subset of $D$.
   \item[(j)] $W \cup X \cup Y = \bigcup\limits_{k=1}^m \partial V_k$.
   \item[(k)] $Y = \!\bigcup\limits_{1 \le k < l \le m}
                        (Y_{kl} \cap \partial V_k \cap \partial V_l)$.
   \item[(l)]  $Y$ is dense-in-itself.
\end{itemize}
\end{lemma}

\proof
(a)--(c) are trivial.
%% (a): Here's the proof that $C$ closed --> $\partial C$ has empty interior.
%%    Suppose $U$ open $\subset \partial C \subset C$.
%%    Then $U \in C^\circ = C \setminus \partial C$,
%%      hence $U = \emptyset$.

(d) Let $z_0 \in Y_{kl} \subset D_k \cap D_l$,
so that $|\beta_k(z_0)| = |\beta_l(z_0)| > 0$.
Then $\beta_k/\beta_l$ is analytic in a neighborhood $U \ni z_0$.
If $\beta_k/\beta_l$ equals a constant $\omega$,
then the no-degenerate-dominance condition implies that $D_k$ ($=D_l$)
must have empty interior, so trivially $z_0 \in \partial D_k = \partial D_l$.
If, on the other hand, $\beta_k/\beta_l$ is nonconstant,
then the open mapping theorem implies that $\beta_k/\beta_l$ must
take values both inside and outside the unit circle in every neighborhood
of $z_0$.  The former points belong to $D_k^c$, and the latter to $D_l^c$;
so $z_0 \in \partial D_k \cap \partial D_l$.

(e) is trivial.

(f) $V_k \cup W_k$ is clearly open, hence contained in $D_k^\circ$.
On the other hand, by (d), $D_k^\circ \cap Y_{kl} = \emptyset$
for each $l \neq k$, so $D_k^\circ \subset V_k \cup W_k \cup X$.

(g) and (h) are trivial consequences of the assumption that
none of the $\alpha_k$ or $\beta_k$ are identically zero.

(i) We use the fact (valid in arbitrary topological spaces
\cite[Exercise 1.3.D]{Engelking_77})
that if $A$ is closed, then for every set $B$ we have
$(A \cup B)^\circ = (A \cup B^\circ)^\circ$.
In particular, if $A$ and $B$ have empty interior
and at least one of them is closed, then $A \cup B$ has empty interior.
Applying this to $A=X$ and $B=Y_{kl}$,
we conclude from (d) and (h) that $X \cup Y_{kl}$ has empty interior.
Moreover, by (c), $X \cup Y_{kl}$ is closed.
Further repeated applications, using (g),
then lead to the conclusion that $W \cup X \cup Y$
is a closed set with empty interior.

(j) By (i), $V$ is open and dense, so $W \cup X \cup Y = V^c = \partial V$.
On the other hand,
$\partial\Biggl( \bigcup\limits_{k=1}^m V_k \Biggr)
   = \bigcup\limits_{k=1}^m \partial V_k$
for any finite collection of disjoint open sets in any topological space
\cite[Theorem 1.3.2(iii) and Exercise 1.3.A]{Engelking_77}.
%%
%% Thanks to Norman Weiss for sketching the proof of (j).
%%

(k) Let $z_0 \in Y$, and let $S$ be the set of indices that are dominant
at $z_0$.  Let $U$ be an open neighborhood of $z_0$, chosen small enough
so that no index in $S^c$ is dominant at any point of $U$.
By (i), $U \cap V$ is dense in $U$.
I claim that there must exist at least two distinct indices $k \in S$
for which $U \cap V_k$ is nonempty.
For suppose that there is only one such index,
so that $U \cap V_k$ is dense in $U$.
Then $U \subset D_k$ and $z_0 \in Y_{kl}$ for some $l \neq k$ ($l \in S$).
Now, shrinking $U$ if necessary so that $\beta_k$ is nonvanishing in $U$,
we have $|\beta_l/\beta_k| \le 1$ in $U$,
with $|\beta_l(z_0)/\beta_k(z_0)| = 1$;
so by the maximum modulus theorem we must have $\beta_l \equiv \omega \beta_k$
for some constant $\omega$ with $|\omega| = 1$,
contrary to the no-degenerate-dominance condition.
Hence there are indices $k \neq l$ ($k,l \in S$)
such that $U \cap V_k$ and $U \cap V_l$ are both nonempty.
Since this holds for arbitrarily small $U \ni z_0$,
and since there are finitely many pairs $k,l$,
%%  {\bf Can I find a cleaner way to do this latter argument????}
we must have $z_0 \in Y_{kl} \cap \partial V_k \cap \partial V_l$
for some pair $k \neq l$.

(l)   Let $z_0 \in Y$, and let $U$ be any connected open neighborhood of $z_0$,
chosen small enough so that $U \cap (W \cup X) = \emptyset$;
and let $U' = U \setminus \{z_0\}$.
If it were true that $U' \cap Y = \emptyset$,
then we would have $U' \subset \bigcup_{k=1}^m V_k$.
But since the $V_k$ are open and disjoint, and $U'$ is connected,
$U'$ would have to be contained in one set $V_k$;
but this contradicts the fact (k) that every point in $Y$
has nearby points in at least two sets $V_k$ and $V_l$.
%%
%% Thanks to Norman Weiss for sketching the proof of (l).
%%

\qed

\noindent
{\bf Remarks.}
1.  It is not necessarily true that
$Y_{kl} \subset \partial V_k \cap \partial V_l$.
For example, let $D = \C \setminus \{0\}$ and $m=3$,
and let $\beta_1(z) = z$, $\beta_2(z) = 1/z$, $\beta_3(z) = 1$, 
$\alpha_1(z) = \alpha_2(z) = \alpha_3(z) = 1$.
Then $Y_{12} = Y_{13} = Y_{23} = \partial V_1 = \partial V_2 =$ unit circle,
but $V_3 = \emptyset$.

2.  $Y$ cannot have isolated points,
but the individual sets $Y_{kl}$ can:
though in a neighborhood of $z_0 \in Y_{kl}$
there certainly exist nearby points where $|\beta_k| = |\beta_l|$,
such points may fail to belong to $Y_{kl}$ because the indices $k,l$
fail to remain dominant.
{\em Example:}\/  $\beta_1(z) = 1+z$, $\beta_2(z) = 1-z$, $\beta_3(z) = 1$.
Then $Y_{12}$ = imaginary axis, but $Y_{13} = Y_{23} = \{0\}$.

3.  In this paper we won't use parts (j) and (l) of Lemma~\ref{lemma3.8},
but they are useful facts to know in applications.

\bigskip

Let us now define
\be
   u_n(z)   \;=\; {1 \over n} \log |f_n(z)|   \;=\; {1 \over n} \Re \log f_n(z)
   \;,
\ee
which is well-defined everywhere on $D$ provided that we give it
the value $-\infty$ at the zeros of $f_n$.
Clearly, $u_n$ is a continuous map from $D$ into $\R \cup \{-\infty\}$,
and is a harmonic function on $D \setminus \scrz(f_n)$.

We can compute $\lim\limits_{n \to\infty} u_n(z)$ at nearly every
point $z \in D$.  Let us define
\begin{eqnarray}
   S(z)          & = &  \{ k \colon\; \alpha_k(z) \neq 0 \}    \\[2mm]
   \betatilde(z) & = &  \cases{ \max \{|\beta_k(z)| \colon\; k \in S(z) \}
                                    & if $S(z) \neq \emptyset$  \cr
                                \noalign{\vskip 2mm}
                                0   & if $S(z) = \emptyset$  \cr
                              }
         \\[2mm]
   T(z)   & = &  \{ k \in S(z) \colon\;  |\beta_k(z)| = \betatilde(z) \}
\end{eqnarray}
It then follows easily from the definition of $f_n$ that:

\begin{lemma}
   \label{lemma3.9}
Under the hypotheses of Theorem~\ref{thm1.5}:
\begin{itemize}
   \item[(a)]  For $z \in V_k$,
      $\lim\limits_{n \to\infty} u_n(z) = \log |\beta_k(z)| > -\infty$,
      and the convergence is uniform on compact subsets of $V_k$.
   \item[(b)]  More generally, if $T(z) = \{k\}$, then
      $\lim\limits_{n \to\infty} u_n(z) = \log |\beta_k(z)|$,
      though this may equal $-\infty$.
   \item[(c)]  For $z \in X$, $u_n(z) = -\infty$ for all $n$,
      so that $\lim\limits_{n \to\infty} u_n(z) = -\infty$.
      More generally, this occurs whenever $\betatilde(z) = 0$
      [i.e.\ when $\alpha_k(z) \beta_k(z) = 0$ for all $k$].
   \item[(d)]  For all $z \in D$,
      $\limsup\limits_{n \to\infty} u_n(z) \le \log \betatilde(z)$.
      In particular, for $z \in W_k$,
      $\limsup\limits_{n \to\infty} u_n(z) < \log |\beta_k(z)|$.
\end{itemize}
\end{lemma}

\medskip
\noindent
{\bf Remark.}
The behavior of the sequence $\{ u_n(z) \}$ is subtle at points $z$
where two or more dominant indices compete, i.e.\ where $|T(z)| \ge 2$
and $\betatilde(z) > 0$.
But we won't need this information.

\proofof{Theorem~\ref{thm1.5}}
If $z_0 \in V_k$, then choose $\epsilon > 0$ so that
$\overline{D}_\epsilon \equiv \{ z \colon\; |z-z_0| \le \epsilon \}
 \subset V_k$;
it follows that there exist constants $\delta > 0$ and $M < \infty$ such that
\begin{subeqnarray}
   |\alpha_k(z)|  & \ge &   \delta   \\[2mm]
   \left| {\alpha_l(z) \over \alpha_k(z)} \right|  & \le &  M
       \qquad\hbox{for all } l \neq k   \\[2mm]
   |\beta_k(z)|   & \ge &   \delta   \\[2mm]
   \left| {\beta_l(z) \over \beta_k(z)} \right|  & \le &  1 - \delta
       \qquad\hbox{for all } l \neq k
\end{subeqnarray}
for all $z \in \overline{D}_\epsilon$.  Then
\begin{subeqnarray}
   |f_n(z)|   & \ge &   |\alpha_k(z)| \, |\beta_k(z)|^n   \;-\;
             \sum\limits_{l \colon\, l \neq k} |\alpha_l(z)| \, |\beta_l(z)|^n
      \\[2mm]
   & \ge &  \delta^{n+1} \left[ 1 \,-\, (m-1) M (1-\delta)^n \right]
\end{subeqnarray}
for all $z \in \overline{D}_\epsilon$.
Therefore, $f_n$ is nonvanishing on $\overline{D}_\epsilon$
for all sufficiently large $n$, so that $z_0 \notin \limsup \scrz(f_n)$.

If $z_0 \in W_k$, we have
$\limsup\limits_{n \to\infty} u_n(z_0) \le \log \betatilde(z_0)
   < \log |\beta_k(z_0)|$ by Lemma~\ref{lemma3.9}(d).
On the other hand, by Lemmas~\ref{lemma3.8}(g) and \ref{lemma3.9}(a)
there exists a punctured neighborhood of $z_0$ on which
$\lim\limits_{n \to\infty} u_n(z) = \log |\beta_k(z)|$;
and this quantity tends to $\log |\beta_k(z_0)|$ as $z \to z_0$.
Therefore, there does not exist a neighborhood $U \ni z_0$
and a continuous function $v \colon\; U \to \R \cup \{-\infty\}$
satisfying $\liminf\limits_{n\to\infty} u_n(z) \le v(z)
 \le \limsup\limits_{n\to\infty} u_n(z)$ for all $z \in U$.
In particular, we can apply Theorem~\ref{thm3.2}
and conclude that $z_0 \in \liminf \scrz(f_n)$.

If $z_0 \in X$, we have $f_n(z_0) = 0$ for all $n$,
so trivially $z_0 \in \liminf \scrz(f_n)$.

Finally, if $z_0 \in Y$, Lemma~\ref{lemma3.8}(k) implies that
there exist indices $k \neq l$ such that
$z_0 \in Y_{kl} \cap \partial V_k \cap \partial V_l$.
So let $U$ be any open neighborhood of $z_0$,
chosen small enough so that $\beta_k$ and $\beta_l$ are
nonvanishing on $U$;
then the sets $U_k \equiv U \cap V_k$ and $U_l \equiv U \cap V_l$
are nonempty.
Lemma~\ref{lemma3.9}(a) implies that
\be
   \lim\limits_{n \to\infty} u_n(z)  \;=\;
   \cases{ \log |\beta_k(z)|   & for $z \in U_k$   \cr
           \log |\beta_l(z)|   & for $z \in U_l$   \cr
         }
\ee
Now there cannot exist a harmonic function $v$ on $U$
that coincides with $\lim\limits_{n \to\infty} u_n(z)$ on $U_k \cup U_l$:
for if there did, then by the uniqueness of harmonic continuation
we would have $|\beta_k(z)| = |\beta_l(z)| = e^{v(z)}$ on $U$,
hence $\beta_k \equiv \omega \beta_l$ for some constant $\omega$
with $|\omega| = 1$, contrary to the no-degenerate-dominance condition.
Therefore, by Theorem~\ref{thm3.2}, we have
$z_0 \in \liminf \scrz(f_n)$.
\qed

\section{Proof of a Strong Form of Lemma~\protect\ref{lemma1.6}}  \label{sec4}

I will prove the following strengthened version of Lemma~\ref{lemma1.6}:

\begin{theorem}
   \label{thm4.1}
Let $F_1,F_2,G$ be analytic functions on a domain $D \subset \C$,
with $G \not\equiv \hbox{constant}$.  For each integer $s \ge 0$, define
\be
   \scru_s   \;=\;
   \{ z \in D \colon\; |1 + F_1(z) G(z)^s|   \,=\,  |1 + F_2(z) G(z)^s| \}
   \;.
\ee
Then
$
  \liminf \scru_s  \;=\;  \limsup \scru_s  \;=\;
      \{ z \in D \colon\; |G(z)| \le 1 \} \,\bigcup\,
      \{ z \in D \colon\; |F_1(z)| = |F_2(z)| \}
  \;.
$
\end{theorem}

In the proof of Theorem~\ref{thm4.1},
I will need the following elementary consequence of
Taylor's theorem with remainder:

\begin{lemma}
 \label{lemma4.2}
Let $w$ be a complex number.  Then $|\log(1+w) - w| \le |w|^2$
provided that $|w| \le 1/4$, where ``log'' denotes the principal branch.
\end{lemma}

\bigskip\par\noindent
{\sc Proof of Theorem~\ref{thm4.1}.\ }
It is easy to see that $\limsup \scru_s \subset 
   \{ z \in D \colon\; |G(z)| \le 1 \} \,\cup\,
   \{ z \in D \colon\; |F_1(z)| = |F_2(z)| \}$:
for if $z_0 \in D$ with $|G(z_0)| > 1$ and $|F_1(z_0)| > |F_2(z_0)|$ (say),
then we can find a neighborhood $U \ni z_0$ on which
$|G(z)| \ge 1+\delta$ and $|F_1(z)| - |F_2(z)| \ge \delta$,
for some $\delta > 0$;
but then we have $|1 + F_1(z) G(z)^s|  >  |1 + F_2(z) G(z)^s|$
for all $s$ large enough so that $\delta (1+\delta)^s > 2$.

So it suffices to prove that
(a) $\liminf \scru_s \supset \{ z \in D \colon\; |G(z)| \le 1 \}$
and
(b) $\liminf \scru_s \supset
     \{ z \in D \colon\; |G(z)| > 1 \hbox{ and } |F_1(z)| = |F_2(z)| \}$.

(a)  Assume that $z_0 \in D$ with $|G(z_0)| \le 1$.
Fix any $\epsilon > 0$ small enough so that
$D_\epsilon \equiv \{z\colon\; |z - z_0| < \epsilon\} \subset D$.
Since $G$ is nonconstant, by the open mapping theorem
we can find $z_1 \in D_\epsilon$ such that
$|G(z_1)| < 1$ and $G'(z_1) \neq 0$;
and we can then find an open set $U \ni z_1$
with $U \subset \overline{U} \subset D_\epsilon$,
and a number $\rho < 1$,
such that $G$ is a homeomorphism of $U$ onto
$G[U] \subset \{w\colon\; |w| < \rho\}$.
%%
%% Thanks to Norman Weiss for suggesting to restrict the domain so that
%%   G is a homeomorphism, thereby allowing a simpler definition of C below.
%%
If $F_1 \equiv F_2$, the claim we want to prove is trivial;
so assume $F_1 \not\equiv F_2$ and choose an open set $V \subset U$
such that
\be
   (F_1 - F_2)[V] \;\subset\;
   \{w\colon\; |w| \ge \delta \hbox{ and }
       \varphi_1 - {\pi \over 4} \le \arg w \le \varphi_1 + {\pi \over 4} \}
\ee
for some $\delta > 0$ and some $\varphi_1$.
Since $\overline{V} \subset D_\epsilon$, we have
\be
   M  \;\equiv\;  \sup\limits_{z \in V}  [|F_1(z)| + |F_2(z)|]  \;<\; \infty
   \;.
\ee

Since $G[V]$ is an open subset of $\{w\colon\; |w| < \rho\}$,
it contains an arc
$A_{r,\theta_1,\theta_2} \equiv
 \{w=re^{i\theta} \colon\; \theta_1 \le \theta \le \theta_2 \}$
for some $0 < r < \rho$ and $\theta_1 < \theta_2$.
Let $C = U \cap G^{-1}[A_{r,\theta_1,\theta_2}]$;
since $G \restrict U$ is a homeomorphism,
we know that $C$ is a topological arc (and in particular a connected set)
contained in $V$, satisfying $G[C] = A_{r,\theta_1,\theta_2}$.
Now take $s_1 = 2\pi/(\theta_2 - \theta_1)$.
It follows that for each integer $s \ge s_1$,
the set $G^s[C] \equiv \{G(z)^s \colon\; z \in C \}$
is exactly the circle $\{w\colon\; |w| = r^s\}$.

Now choose $s_2$ so that $M \rho^{s_2} \le 1/4$.
Then, for any integer $s \ge s_2$, the function
\be
   L_s(z) \;=\; \log[1 + F_1(z) G(z)^s] \,-\, \log[1 + F_2(z) G(z)^s]
\ee
(where ``log'' denotes the principal branch) is analytic on $V$.
Furthermore, by Lemma~\ref{lemma4.2} we have
\be
   \left| L_s(z) \,-\, [F_1(z) - F_2(z)] G(z)^s \right|
   \;\le\;  2 (M r^s)^2
\ee
for $z \in C$.

For each integer $s \ge s_1$, there exist points $z_s^+, z_s^- \in C$
such that $G(z_s^\pm)^s = \pm r^s e^{-i \varphi_1}$ and hence
\be
   \pm \real \left\{ [F_1(z_s^\pm) - F_2(z_s^\pm)] \, G(z_s^\pm)^s \right\}
   \;\ge\;  {\delta r^s \over \sqrt{2}}
   \;.
\ee
For $s \ge \max(s_1,s_2)$ we then have
\be
   \pm \real L_s(z_s^\pm)   \;\ge\;
   {\delta r^s \over \sqrt{2}} \,-\,  2 M^2 r^{2s}
   \;.
\ee
Now choose $s_0 \ge \max(s_1,s_2)$ large enough so that
$2 M^2 r^{s_0} < \delta/\sqrt{2}$.
It follows that for $s \ge s_0$ we have $\pm \real L_s(z_s^\pm) > 0$.
Since $L_s$ is continuous and $C$ is connected,
it follows that there exists $z_s^{(0)} \in C$ such that
$\real L_s(z_s^{(0)}) = 0$.
This completes the proof of assertion (a).

(b)
Assume that $z_0 \in D$ with $|G(z_0)| > 1$ and $|F_1(z_0)| = |F_2(z_0)|$.
If $F_1(z_0) = F_2(z_0) = 0$, the claim is trivial
(we have $z_0 \in \scru_s$ for all $s$);
so assume that $|F_1(z_0)| = |F_2(z_0)| > 0$.
Fix any $\epsilon > 0$ small enough so that
$D_\epsilon \equiv \{z\colon\; |z-z_0| < \epsilon\} \subset D$
and
\begin{subeqnarray}
   |G(z)|    & \ge &  1+\delta \\
   |F_1(z)|  & \ge &  \delta \\
   |F_2(z)|  & \ge &  \delta
\end{subeqnarray}
for all $z \in D_\epsilon$, for some $\delta > 0$.
We then consider two cases:

{\em Case 1: $F_1 \equiv \omega F_2$ for some $\omega \in \C$}\/
(obviously $|\omega| = 1$).
We need to show that for all $s$ sufficiently large,
there exists $z \in D_\epsilon$ such that
\be
   | 1 + F_1(z)^{-1} G(z)^{-s} |   \;=\;
   | 1 + F_2(z)^{-1} G(z)^{-s} |   \;.
\ee
But this follows from part (a) applied to the functions
$1/F_1$, $1/F_2$ and $1/G$.

{\em Case 2: There does not exist $\omega \in \C$ such that
   $F_1 \equiv \omega F_2$.}\/
In this case the function
\be
   f(z)  \;=\;  F_1(z_0) F_2(z) \,-\, F_2(z_0) F_1(z)
\ee
has a zero at $z=z_0$ but is not identically vanishing.
So it follows from Rouch\'e's theorem that for all sufficiently large $s$,
the equation
\be
   F_1(z_0) F_2(z) \,-\, F_2(z_0) F_1(z) \,+\,
    [F_1(z_0) - F_2(z_0)] G(z)^{-s}   \;=\;  0
\ee
has a solution $z \in D_\epsilon$.  But this means that
\be
   {F_1(z) + G(z)^{-s}  \over  F_1(z_0)}  \;=\;
   {F_2(z) + G(z)^{-s}  \over  F_2(z_0)}  \;,
\ee
hence that $z \in \scru_s$.
\qed

\section{Completion of the Proof of Theorem~\ref{thm1.2bis}}  \label{sec5}

We can now put together all the ingredients to complete the proof of
Theorem~\ref{thm1.2bis} (and hence also of the weaker Theorem~\ref{thm1.2})
along the lines sketched in Section~\ref{sec1.3}.
We shall give the details for part (a);
the proof of (b) is almost identical, {\em mutatis mutandis}\/.

Fix a complex number $v_0 \neq 0$, and fix $\epsilon > 0$ and $R < \infty$.
Since the open region
$D \equiv \{ q \in \C \colon\; |q + v_0| > |v_0| \hbox{ and } |q| < R\}$
has compact closure
$\bar{D} \equiv
 \{ q \in \C \colon\; |q + v_0| \ge |v_0| \hbox{ and } |q| \le R\}$,
we can choose finitely many points $q_1,\ldots,q_n \in D$
such that each point $q \in \bar{D}$ lies within a distance
$\epsilon/3$ of at least one $q_i$.

Now, for each $i$ ($1 \le i \le n$),
apply Lemma~\ref{lemma1.6} (or the stronger Theorem~\ref{thm4.1})
with $z = q-q_i$, $F_1(z) = q-1$, $F_2(z) = -1$, $G(z) = v_0/(q+v_0)$.
[The hypothesis $v_0 \neq 0$ guarantees that $G \not\equiv \hbox{constant}$,
 and the hypothesis $q_i \in D$ guarantees that $|G(0)| < 1$.]
We conclude that, for each $i$, there exists an integer $s_{0,i} < \infty$
such that for all integers $s \ge s_{0,i}$ the equation
\be
   \left| 1 \,+\, (q-1) \left({v_0 \over q+v_0}\right) ^{\! s} \right|
   \;=\;
   \left| 1 \,-\, \left({v_0 \over q+v_0}\right) ^{\! s} \right|
   \;.
  \label{fixed_s_curve_bis}
\ee
[cf.\ \reff{fixed_s_curve}] has a solution in the disc $|q-q_i| < \epsilon/3$.
Let $s_0 = \max\limits_{1 \le i \le n} s_{0,i}$.

Now fix any $s \ge s_0$,
and let $q'_1,\ldots,q'_n$ be solutions of \reff{fixed_s_curve_bis}
satisfying $|q'_i-q_i| < \epsilon/3$ and $q'_i \neq 0$ for all $i$.
Apply Theorem~\ref{thm1.5} (or the original Beraha--Kahane--Weiss theorem)
with $m=2$, $n=p$, $z=q$, $\alpha_1(z) = 1$, $\alpha_2(z) = q-1$,
$\beta_1(z) = (q+v_0)^s + {(q-1)v_0^s}$, $\beta_1(z) = (q+v_0)^s - v_0^s$.
[The nondegeneracy condition holds because $v_0 \neq 0$.]
We conclude that the zeros of
\be
   q^{p-1} Z_{\Theta^{(s,p)}}(q,v)   \;=\;
   [(q+v_0)^s + (q-1)v_0^s]^p \,+\, (q-1) [(q+v_0)^s - v_0^s]^p
 \label{Z_gen_theta_numerator}
\ee
[cf.\ \reff{Z_gen_theta}] accumulate (in lim inf as well as lim sup sense)
at each point of the solution set of \reff{fixed_s_curve_bis}.
In particular, for each $i$,
there exists a zero of \reff{Z_gen_theta_numerator}
in the disc $|q-q'_i| < \min(\epsilon/3, |q'_i|)$.
[Since this disc does not contain 0, the given $q$ must be a zero of
 $Z_{\Theta^{(s,p)}}(q,v_0)$.]
Set $p_0 = \max\limits_{1 \le i \le n} p_{0,i}$.

We have shown that for each $p \ge p_0$,
there exist zeros $q''_1,\ldots,q''_n$ of $Z_{\Theta^{(s,p)}}(\,\cdot\,,v_0)$
such that each point $q \in \bar{D}$
lies within distance $\epsilon$ of at least one $q''_i$.
This completes the proof.

\section{Real Chromatic Roots of the Graphs $\Theta^{(s,p)}$}   \label{sec6}

Though this paper is primarily concerned with the
{\em complex}\/ chromatic roots of the graphs $\Theta^{(s,p)}$
[and more generally, with the complex roots of their
 dichromatic polynomials $Z_{\Theta^{(s,p)}}(q,v)$],
let me digress to discuss their {\em real}\/ chromatic roots.
The chromatic polynomial $P_{\Theta^{(s,p)}}(q)$ is obtained by
specializing \reff{Z_gen_theta} to $v=-1$:
\be
   P_{\Theta^{(s,p)}}(q)   \;=\;
      {[(q-1)^s + (q-1)(-1)^s]^p \,+\, (q-1) [(q-1)^s - (-1)^s]^p
       \over
       q^{p-1}
      }
   \;.
\ee
It is useful to write $y=1-q$, so that
\be
   F(y) \;\equiv\; (-1)^{sp+1} q^{p-1} P_{\Theta^{(s,p)}}(q)   \;=\;
   y (y^s -1)^p  \,-\, (y^s - y)^p \;.
\ee
Recall that $\Theta^{(s,p)}$ has $n \equiv (s-1)p + 2$ vertices.
I write $P^{(j)}_G(q)$ to denote the $j^{th}$ derivative of $P_G(q)$.
The real chromatic roots of the graphs $\Theta^{(s,p)}$
are then fully characterized by following result:

\begin{proposition}
   \label{prop5.1}
Fix integers $s,p \ge 2$ and set $n = (s-1)p+2$.  Then:
\begin{itemize}
   \item[(a)]  $(-1)^{n+j} P^{(j)}_{\Theta^{(s,p)}}(q) > 0$
       for $q<0$ and $0 \le j \le n$.
   \item[(b)]  $P_{\Theta^{(s,p)}}(0) = 0$ and
       $(-1)^n P'_{\Theta^{(s,p)}}(0) = (s-1)^p - s^p < 0$.
   \item[(c)]  $(-1)^n P_{\Theta^{(s,p)}}(q) < 0$ for $0 < q < 1$.
   \item[(d)]  $P_{\Theta^{(s,p)}}(1) = 0$ and
       $(-1)^n P'_{\Theta^{(s,p)}}(1) = 1 > 0$.
   \item[(e)]  If $n$ is even (i.e.\ $s$ is odd or $p$ is even
       or both), then $P_{\Theta^{(s,p)}}(q) > 0$
       for $q > 1$.
   \item[(e$\,{}'$)]  If $n$ is odd (i.e.\ $s$ is even and $p$ is odd),
       then there exists $q_*^{(s,p)} \in [1.4301597\ldots,2)$ such that:
\begin{itemize}
   \item[(e$\,{}'_1$)]  $P_{\Theta^{(s,p)}}(q) < 0$
       for $1 < q < q_*^{(s,p)}$;
   \item[(e$\,{}'_2$)]  $P_{\Theta^{(s,p)}}(q_*^{(s,p)}) = 0$ and
       $P'_{\Theta^{(s,p)}}(q_*^{(s,p)}) > 0$;  and
   \item[(e$\,{}'_3$)]  $P_{\Theta^{(s,p)}}(q) > 0$
       and $P'_{\Theta^{(s,p)}}(q) > 0$ for $q > q_*^{(s,p)}$.
\end{itemize}
Here $1.4301597\ldots$ ($= q_*^{(2,3)}$) is shorthand for the
unique real root of $q^3 - 5q^2 + 10q - 7 = 0$.
Moreover,
\begin{itemize}
   \item[(e$\,{}'_{\rm a}$)]  $q_*^{(s,p)}$ is a strictly increasing function
       of $s$ and $p$;
   \item[(e$\,{}'_{\rm b}$)]  $\lim\limits_{s \to\infty} q_*^{(s,p)} = 2$
       for each $p$; and
   \item[(e$\,{}'_{\rm c}$)]  $\lim\limits_{p \to\infty} q_*^{(s,p)} =
       q_*^{(s,\infty)}$, where $q_*^{(s,\infty)} \in [{3 \over 2}, 2)$
       is the unique nonzero real root of $2(q-1)^s + q-2 = 0$.
\end{itemize}
\end{itemize}
\end{proposition}

\noindent
The signs of $P_G(q)$ and its derivatives in (a)--(d)
are of course general facts
about the chromatic polynomial of {\em any}\/ 2-connected graph
\cite[Theorem 2]{Woodall_77},
but it is amusing to verify them explicitly for $\Theta^{(s,p)}$.

\medskip

\proof
(a) We have
\begin{subeqnarray}
   F(y)   & = &   (y-1)(y^s -1)^p  \,+\, [(y^s -1)^p - (y^s -y)^p]  \\[2mm]
          & = &   (y-1) \left[ (y^s -1)^p  \,+\,
                  \sum\limits_{k=0}^{p-1}  (y^s -1)^k (y^s -y)^{p-1-k}
                        \right]   \;,
\end{subeqnarray}
so clearly $F$ and all its derivatives are nonnegative for $y \ge 1$.
If we now write $P_{\Theta^{(s,p)}}(q) = \sum_{k=1}^n a_k^{(s,p)} q^k$,
we see that $(-1)^{n-k} a_k^{(s,p)} = F^{(k+p-1)}(1) / (k+p-1)!  \ge 0$;
and of course $a_n^{(s,p)} = 1 > 0$.  The claims follow.

(b) An easy calculation.

(c) For $0 < y < 1$, we have $(1-y^s)/(y-y^s) > 1/y > 1$
and hence $[(1-y^s)/(y-y^s)]^p > 1/y$,
so that $(-1)^p F(y) > 0$.

(d) An easy calculation.

(e) For $y < 0$ and $p$ even, we manifestly have $F(y) < 0$.
For $y = -z < 0$ and $s$ odd, we have
$(-1)^{p+1} F(y) = z (z^s +1)^p + (z^s -z)^p$.
For $z \ge 1$ this is manifestly $>0$;
for $0 < z < 1$ it is $> z - z^p > 0$ as well.
 
(e\textprime)  For $z = q-1 > 0$ with $s$ even and $p$ odd, we can write
$P_{\Theta^{(s,p)}}(q)  = (1+z) G(z)^p [1 - H(z)^p]$
where
\begin{subeqnarray}
   G(z)   & = &   {z + z^s  \over  1+z}   \\[2mm]
   H(z)   & = &   {1-z^s \over z^{(p-1)/p} (1+ z^{s-1})}
\end{subeqnarray}
Now $H(z)$ is a strictly decreasing function of $z$ on $(0,\infty)$,
which runs from $+\infty$ at $z=0$ to 0 at $z=1$ to $-\infty$ at $z=+\infty$;
therefore, the equation $H(z) = 1$ has a unique solution
$z_*^{(s,p)} \in (0,\infty)$, which lies in fact in $(0,1)$.
Moreover, $H(z)$ is a strictly increasing function of $s$ and $p$
when $z \in (0,1)$,
so $z_*^{(s,p)}$ is a strictly increasing function of $s$ and $p$;
and it is easy to see that it has properties
(e\textprime${}_{\rm b}$) and (e\textprime${}_{\rm c}$).
Finally, $G(z)$ is a strictly increasing function of $z$ on $(0,\infty)$,
so $P'_{\Theta^{(s,p)}}(q) > 0$ for $q \ge q_*^{(s,p)}$.
\qed

\noindent
{\bf Remarks.}
1.  For any $n$-vertex connected graph $G$ and any vertex $v$ of $G$,
the quantity $T_G(1,0) = (-1)^{n+1} P'_G(0)$ counts the acyclic orientations
of $G$ in which $v$ is the unique source
\cite[Theorem 7.3]{Greene_83}  \cite[Proposition 6.3.18]{Brylawski_92}
\cite{Gebhard_00}.
Taking $v$ to be one of the endvertices of $\Theta_{s_1,\ldots,s_p}$,
it is easy to show that this quantity equals
$\prod\limits_{i=1}^p s_i - \prod\limits_{i=1}^p (s_i-1)$,
exactly as computed from the chromatic polynomial.
%% PROOF:  The edges incident on $v$ must all be directed outwards.
%%   On path $i$, we can let the last $k_i$ edges ($0 \le k_i \le s_i -1$)
%%   point backwards (i.e. towards $v$).  But we must not take $k_i \ge 1$
%%   for ALL $i$, for this would make the other endvertex into a source.

2.  For any $n$-vertex connected graph $G \neq K_2$
and any edge $e = \<uv\>$ of $G$,
the quantity $[\partial T_G(x,y)/\partial x](0,0) = (-1)^n P'_G(1)$
counts the acyclic orientations of $G$
in which $u$ is the unique source and $v$ is the unique sink
\cite[Theorem 7.2]{Greene_83}  \cite[Exercise 6.35]{Brylawski_92}
\cite{Gebhard_00}.
It also equals half the number of totally cyclic orientations of $G$
(i.e.\ orientations in which every edge of $G$ belongs to some
 directed cycle)
in which every directed cycle uses $e$
\cite[Theorem 8.2]{Greene_83}
\cite[Proposition 6.2.12 and Example 6.3.29]{Brylawski_92}.
Taking $u$ to be one of the endvertices of $\Theta_{s_1,\ldots,s_p}$
and $v$ one of its neighbors,
it is easy to see that there is exactly one orientation of the first kind
(and it is acyclic), and two of the second.

3.  In part (e) we {\em cannot}\/ assert that
$P_{\Theta^{(s,p)}}(q)$ is increasing for $q>1$.
Indeed, for many pairs $(s,p)$ --- including
$(2,4)$, $(2,6)$, $(4,6)$, $(5,6)$, $(5,7)$, $\ldots$ ---
there is an interval $1 < q_1 < q < q_2 < 2$
where $P'_{\Theta^{(s,p)}}(q) < 0$.

4.  In part (e\textprime), $P_{\Theta^{(s,p)}}(q)$ is not necessarily
convex on $(1,2)$:  for example, for $(s,p) = (2,7)$
there are inflection points at $q \approx 1.282916, 1.405642$.
However, numerical calculations for small $s,p$ suggest that
$P_{\Theta^{(s,p)}}(q)$ is convex on the interval $[q_*^{(s,p)}, \infty)$.
Indeed, it appears that {\em all}\/ the derivatives of $P_{\Theta^{(s,p)}}(q)$
are strictly positive at $q = q_*^{(s,p)}$,
except that for $s=2$ and $p \ge 7$ the $(n-1)^{st}$ derivative is $< 0$.
%%
%% \Theta^{(s,p)} has n = (s-1)p + 2 vertices and m = sp edges.
%% And P_G(q) = q^n - m q^{n-1} + ...
%% So the (n-1)^st derivative is > 0 for q > m/n and is < 0 for q < m/n.
%% For s=2, we have m/n = 2p/(p+2), which is < q_*^{(2,p)} for p=3,5;
%%   but for p \ge 7 it is > 3/2 > q_*^{(2,p)}, so the condition fails.
%% On the other hand, for s \ge 4 we have m/n = sp/[(s-1)p + 2] < s/(s-1) < 4/3,
%%   while q_*^{(s,p)} \ge q_*^{(4,3)} \approx 1.6021;
%%   so at least the (n-1)^st derivative behaves correctly in these cases.
%%
%{\bf In these latter cases, do any derivatives other than the $(n-1)^{st}$
%     become negative somewhere on $[q_*^{(s,p)}, \infty)$????}

\bigskip

Though I am unable to say much about the monotonicity or convexity
of $P_{\Theta^{(s,p)}}(q)$ for $1 < q < 2$
(cf.\ Remarks 3 and 4 above),
the situation simplifies dramatically for $q \ge 2$
for all series-parallel graphs.
We begin with the following:

\begin{definition}
 \label{def_strongly_positive}
Let $P$ be a degree-$n$ polynomial with real coefficients,
and let $a \in \R$.
We say that $P$ is {\em strongly nonnegative}\/
(resp.\ {\em strongly positive}\/)
at $a$ in case $P^{(j)}(a) \ge 0$ (resp.\ $P^{(j)}(a) > 0$)
for all $0 \le j \le n$.
\end{definition}

\noindent
Note that if $P$ is strongly nonnegative at $a$ and is not identically zero,
then $P$ is strongly positive on $(a,\infty)$;
and conversely, if $P$ is strongly nonnegative on $(a,\infty)$,
then it is also strongly nonnegative at $a$.
Note also that if $P$ and $Q$ are strongly nonnegative
(resp. strongly positive) at $a$,
then so are $P+Q$ and $PQ$.
I suspect that strong positivity is the appropriate concept
for many (though perhaps not all) theorems asserting
upper zero-free intervals for chromatic polynomials
\cite{Birkhoff_46,Woodall_97,Thomassen_97}.
Here, at any rate, is one example:

\begin{proposition}
   \label{prop5.3}
Let $G=(V,E)$ be a finite graph equipped with edge weights
$\{ v_e \} _{e \in E}$;
let $x,y$ be distinct vertices of $G$;
and let
\be
   Z_{G,x,y}(q, \{v_e\}; \sigma_x, \sigma_y)   \;=\;
   A_{G,x,y}(q, \{v_e\})  \,+\,
      B_{G,x,y}(q, \{v_e\}) \, \delta(\sigma_x,\sigma_y)
\ee
be its restricted partition function.
Suppose that the two-terminal graph $(G,x,y)$ is series-parallel\/\footnote{
   Note that this is {\em stronger}\/ than requiring simply that
   $G$ be series-parallel;
   we require that $G$ {\em with the two fixed terminals $x$ and $y$}\/
   be transformable to a single edge by a sequence of
   series and parallel reductions.
},
and that $v_e \ge -1$ for all edges $e \in E$.
Then $A$ and $A+B$ are strongly nonnegative at $q=2$.
\end{proposition}

\begin{corollary}
   \label{cor5.4}
Let $G$ be a series-parallel graph equipped with edge weights
$\{ v_e \} _{e \in E}$ satisfying $v_e \ge -1$ for all edges $e$.
Then $Z_G(q,\{v_e\})/q$ [and hence also $Z_G(q,\{v_e\})$]
is strongly nonnegative at $q=2$.
In particular, $P_G(q) > 0$ for $q > 2$.
\end{corollary}

The proof of Proposition~\ref{prop5.3} is an easy application
of the formulae \reff{eq2.parallel2} and \reff{eq2.seriesAB}
for parallel and series connection, which can be rewritten as
\begin{eqnarray}
   \biggl( (A_1,B_1), \, (A_2,B_2) \biggr)
   & \stackrel{\rm parallel}{\mapsto} &
   (A^{\rm par}, B^{\rm par}) \,\equiv\,
      (A_1 A_2, A_1 B_2 + A_2 B_1 + B_1 B_2)  \qquad    \\[2mm]
   \biggl( (A_1,B_1), \, (A_2,B_2) \biggr)
   & \stackrel{\rm series}{\mapsto} &
   (A^{\rm ser}, B^{\rm ser}) \,\equiv\,
      (A_1 B_2 + A_2 B_1 + q A_1 A_2, B_1 B_2)  \qquad
\end{eqnarray}

\begin{lemma}
If $A_1$, $A_1 + B_1$, $A_2$ and $A_2 + B_2$ are all strongly nonnegative
at $q=2$, then so are
$A^{\rm par}$, $A^{\rm par} + B^{\rm par}$,
$A^{\rm ser}$ and $A^{\rm ser} + B^{\rm ser}$.
\end{lemma}

\proof
A trivial computation:
\begin{subeqnarray}
   A^{\rm par}    & = &   A_1 A_2   \\[2mm]
   A^{\rm par} + B^{\rm par}   & = &   (A_1 + B_1) (A_2 + B_2)   \\[2mm]
   A^{\rm ser}   & = &  A_1 (A_2 + B_2) + A_2 (A_1 + B_1) + (q-2) A_1 A_2
      \\[2mm]
   A^{\rm ser} + B^{\rm ser}   & = &  (A_1 + B_1) (A_2 + B_2) + (q-1) A_1 A_2
\end{subeqnarray}
\qed

\noindent
Proposition~\ref{prop5.3} then follows by induction,
starting from the fact that it holds when $G$ is a single edge $\< xy \>$
(for which $A=1$ and $A+B = 1 + v_e \ge 0$).
Corollary~\ref{cor5.4} for 2-connected series-parallel graphs $G$
then follows from Proposition~\ref{prop5.3} by choosing some pair of vertices
between which the graph is series-parallel,
and using $Z_G/q = qA + B = (A+B) + (q-1)A$;
it then holds for general series-parallel graphs by factorization
into 2-connected components.

\bigskip

\noindent
{\bf Remark.}
Thomassen \cite[Theorem 3.4]{Thomassen_97} shows that for any integer
$k \ge 2$, the set $M_k$ of real chromatic roots of graphs of
tree-width $\le k$ consists of 0, 1 and a dense subset of the interval
$(32/27, k]$.   Indeed, Thomassen's argument
(which is based on \cite[Theorem 3.3]{Thomassen_97})
actually proves the stronger result that $P_G(q)/q$
[and hence also $P_G(q)$] is strongly nonnegative at $q=k$.
The case $k=2$ corresponds to series-parallel graphs.

\section{Variants of the Construction}   \label{sec7}

\subsection{Adding an edge to $\Theta^{(s,p)}$}   \label{sec7.1}

I recently proved that if $G$ is a graph of maximum degree $\Delta$,
then all the chromatic roots of $G$ lie in the disc $|q| < 7.963907 \Delta$
\cite[Corollary 5.3 and Proposition 5.4]{Sokal_00}.\footnote{
   This result is, in fact, the specialization to chromatic polynomials
   of a more general bound on the zeros of the Potts-model partition function
   $Z_G(\qve)$ throughout the ``complex antiferromagnetic regime'',
   i.e.\ when $|1+v_e| \le 1$ for all edges $e$.
   The other two results of \cite{Sokal_00} cited later in this subsection
   are likewise specializations of bounds valid throughout the
   complex antiferromagnetic regime.  See \cite{Sokal_00} for details.
}
Moreover, if {\em all but one}\/ of the vertices of $G$ have
degree $\le \Delta$, then the chromatic roots of $G$ are still bounded,
namely they lie in the disc $|q| < 7.963907 \Delta + 1$
\cite[Corollary 6.4]{Sokal_00}.
The graphs $\Theta^{(s,p)}$ show that no analogous theorem can hold
if ``all but one'' is replaced by ``all but two''
(not even when $\Delta = 2$ and $G$ is series-parallel),
for in that case the chromatic roots can spread over the whole complex plane
(except perhaps the disc $|q-1| < 1$).

There is at least one instance in which the presence of several vertices
of large degree cannot cause large chromatic roots,
namely when those vertices form an $N$-clique within which
every pair of vertices is connected by a $v=-1$ edge.
In that case, if the remaining vertices of $G$ have degree $\le \Delta$,
then all the chromatic roots of $G$ lie in the disc $|q| < 7.963907 \Delta + N$
\cite[Theorem 6.3(b)]{Sokal_00}.

The condition here that the clique edges have $v=-1$ is crucial.
Indeed, consider the graph $\Theta^{(s,p)} + xy$
(where $x$ and $y$ are the endvertices of $\Theta^{(s,p)}$)
in which the edge $xy$ is assigned a weight $v_{xy}$
and all the other edges have weight $v$.
The corresponding Potts-model partition function is easily calculated from
\reff{eq2.G+xy}, \reff{eq_prop2.1_1} and \reff{eq2.theta}/\reff{eq2.thetaAB}:
\be
   Z_{\Theta^{(s,p)} + xy}(q,v,v_{xy})  \;=\;
   { (1+v_{xy}) [(q+v)^s + (q-1)v^s]^p \,+\, (q-1) [(q+v)^s - v^s]^p
     \over
     q^{p-1}
   }   \;,
\ee
which reduces to \reff{eq2.thetaZqvs} when $v_{xy} = 0$
and to \reff{eq2.thetaZ+xy} when $v_{xy} = -1$.
Provided that $v_{xy} \neq -1$, Theorem~\ref{thm1.5} still applies
and the zeros accumulate as $s,p \to\infty$ in the entire region
where $|v/(q+v)| \le 1$.

\subsection{The dual construction}   \label{sec7.2}

Let $C_n^{(r)}$ be the cycle $C_n$ with each edge replaced by
$r$ edges in parallel.
Note that the dual graph of $\Theta^{(s,p)}$ is $C_p^{(s)}$, and vice versa.
The dichromatic polynomial of $C_n^{(r)}$ is easily computed from
the dichromatic polynomial \reff{eq2.ZCn} of $C_n$
together with the parallel reduction formula \reff{eq2.parallel}:
\be
   Z_{C_n^{(r)}}(q,v)   \;=\;
   [q + (1+v)^r - 1]^n  \,+\,  (q-1) [(1+v)^r - 1]^n
   \;.
\ee
It follows from Theorem~\ref{thm1.5} that when $n \to \infty$
at fixed $r$, the zeros of $Z_{C_n^{(r)}}$ accumulate where
\be
   | 1 \,+\, (q-1) (1+v)^{-r} |  \;=\;  | 1 \,-\, (1+v)^{-r} |
   \;.
\ee
Here the $G \not\equiv \hbox{constant}$ condition of Lemma~\ref{lemma1.6}
means that we cannot fix $v$ and use $q$ as the variable.
However, we can fix $q = q_0 \neq 0$ and use $v$ as the variable\footnote{
   Note that the no-degenerate-dominance condition of
   Theorem~\ref{thm1.5} is satisfied provided that the fixed value of $q$
   is nonzero.
};
we then have:

\begin{theorem}
   \label{thm6.1}
Fix complex numbers $q_0,v_0$ satisfying $q_0 \neq 0$ and $|1+v_0| \ge 1$.
Then, for each $\epsilon > 0$, there exist numbers $r_0 < \infty$
and $n_0(r) < \infty$ such that for all $r \ge r_0$ and $n \ge n_0(r)$,
the dichromatic polynomial $Z_{C_n^{(r)}} (q_0, \,\cdot\,)$
has a zero in the disc $|v - v_0| < \epsilon$.
\end{theorem}

We can also slice the $(q,v)$-space in other ways;
for example, we can fix $u_0 \neq 0$ and take $v = q/u_0$:

\begin{theorem}
   \label{thm6.2}
Fix nonzero complex numbers $q_0, u_0$ satisfying $|u_0| \le |q_0 + u_0|$.
Then, for each $\epsilon > 0$, there exist numbers $r_0 < \infty$
and $n_0(r) < \infty$ such that for all $r \ge r_0$ and $n \ge n_0(r)$,
the polynomial $Z_{C_n^{(r)}} (q, q/u_0)$
has a zero in the disc $|q - q_0| < \epsilon$.

In particular, if $|q_0 - 1| \ge 1$, then the flow polynomial
$F_{C_n^{(r)}}(q) = (-1)^{nr} q^{-n} Z_{C_n^{(r)}} (q, -q)$
has a zero in the disc $|q - q_0| < \epsilon$.
\end{theorem}

Since the $G \not\equiv \hbox{constant}$ condition
forbids us to use $q$ here as the variable with $v$ {\em fixed}\/,
these results are somewhat less interesting
(at least for direct application to chromatic polynomials)
than Theorem~\ref{thm1.2}.
However, they become useful when we insert nontrivial 2-rooted graphs
in place of the edges of $C_n^{(r)}$.
But we defer this construction to a separate paper.

\section{Discussion}  \label{sec8}

It now appears that there are three {\em quite distinct}\/ theories
of chromatic roots:
\begin{itemize}
   \item[(a)]  the theory of {\em integer}\/ chromatic roots
      [i.e.\ the combinatorial theory of the chromatic number $\chi(G)$];
   \item[(b)]  the theory of {\em real}\/ chromatic roots
      (i.e.\ the theory of zero-free and zero-dense intervals for
           various classes of graphs
       \cite{Birkhoff_46,Woodall_77,Woodall_92a,Woodall_92b,%
Jackson_93,Woodall_97,Thomassen_97,Edwards_98,Thomassen_00});  and
   \item[(c)]  the theory of {\em complex}\/ chromatic roots.\footnote{
       There is also a (small) fourth theory,
       namely the ``number-theoretic'' (or ``algebraic'') theory
       of chromatic roots, which exploits the fact that
       the chromatic polynomial is a monic polynomial with {\em integer}\/
       coefficients.
       I include here the proof that the generalized Beraha numbers
       $B_n^{(k)} = 4 \cos^2 (k\pi/n)$ for $n=5,7,8,9$ and $n \ge 11$,
       with $k$ coprime to $n$, are never chromatic roots
       \cite{Salas-Sokal_transfer1}
       (see \cite{Tutte_75a} for earlier related results);
       Tutte's golden-ratio theorems for plane triangulations
       \cite{Tutte_70a,Tutte_70b,Tutte_75a};
       and speculations concerning the accumulation of chromatic roots
       at Beraha numbers $B_n = 4 \cos^2 (\pi/n)$ for certain sequences
       of planar graphs
       \cite{Berman_69,Tutte_74a,Tutte_75a,Tutte_82a,Tutte_82b,%
Baxter_87,Martin_87,Martin_89,Martin_91,%
Saleur_90,Saleur_91,Jaeger_91,Kauffmann_93,Temperley_93,
Tutte_93,Jackson_94,Maillard_94a,Maillard_94b,Maillard_97,%
Salas-Sokal_transfer1,Jacobsen-Salas_transfer2,Jacobsen-Salas-Sokal_transfer3}.
       This theory is, however, as yet rather undeveloped.
}
\end{itemize}

Woodall \cite[Theorem 8]{Woodall_77} has shown that the
complete bipartite graphs $K_{n_1,n_2}$,
in the limit $n_2 \to \infty$ with $n_1$ fixed,
have {\em real}\/ chromatic roots arbitrarily close to all the integers
from 2 through $\lfloor n_1/2 \rfloor$,
even though their only {\em integer}\/ chromatic roots are of course 0 and 1.
More generally, the joins $G = K_{n_1,n_2} + K_n$,
which have chromatic number $\chi(G) = n+2$,
have real chromatic roots arbitrarily close to all the integers
from $n+2$ through $n + \lfloor n_1/2 \rfloor$,
even though their only integer chromatic roots are $0,1,\ldots,n+1$.
This suggests that the real roots of the chromatic polynomial
have nothing much to do with the integer roots.
In particular, the real roots of $P_G(q)$ cannot be bounded
in terms of the chromatic number $\chi(G)$ alone.

Likewise, the results of the present paper show that the
generalized theta graphs $\Theta^{(s,p)}$
have {\em complex}\/ chromatic roots arbitrarily close to any point
in $\C \setminus \{|q-1| < 1\}$,
even though their only {\em real}\/ chromatic roots
are 0, 1 and $q_*^{(s,p)} \in [1.4301597\ldots, 2)$
and their only integer roots are 0 and 1.
This suggests that the complex roots of the chromatic polynomial
have nothing much to do with the real or integer roots.

In particular, planar graphs (such as $\Theta^{(s,p)}$)
can have complex chromatic roots densely surrounding the point 4.\footnote{
   The fact that planar graphs can have chromatic roots arbitrarily close to 4
   was established already two decades ago in the seminal paper of
   Beraha and Kahane \cite{Beraha_79}.  They considered
   $4 \times n$ strips of the triangular lattice with periodic boundary
   conditions in the ``short'' direction
   (and extra sites at both ends of the ``long'' direction,
    but this is inessential)
   and found that the chromatic roots accumulate as $n \to\infty$
   on a curve passing through $q=4$.
}
This suggests that it is unlikely that there will ever be
a ``complex-analysis'' proof of the Four-Color Theorem:
quite simply, $q=4$ behaves very differently from complex values of $q$
arbitrarily close to 4.
On the other hand, it is conceivable that there may some day
be a ``real-analysis'' proof of the Four-Color Theorem
in the same sense that Birkhoff and Lewis \cite[pp.~413--415]{Birkhoff_46}
(see also \cite{Woodall_97,Thomassen_97})
provided a ``real-analysis'' proof of the Five-Color Theorem:
namely, it may be possible to prove that
$P_G(q) > 0$ for real $q \ge 4$, for all loopless planar graphs $G$.
Indeed, Birkhoff and Lewis \cite[p.~413]{Birkhoff_46} have conjectured that
$P_G(q)/[q(q-1)(q-2)] - (q-3)^{n-3}$ is strongly nonnegative
(see Definition~\ref{def_strongly_positive})
at $q=4$ for all loopless planar graphs $G$ with $n$ vertices.

\bigskip

We conclude with a few disconnected remarks:

\medskip

1) Let ${\cal G}_k^<$ denote the class of graphs $G$
such that every subcontraction (= minor) of $G$ has a vertex of
degree $\le k$.
Woodall \cite[Question 1]{Woodall_97} asked whether all the (complex)
chromatic roots of $G \in {\cal G}_k^<$ are bounded in modulus by $k$.
For $k \ge 2$, the answer is no, since
\be
   \hbox{generalized theta graphs} \;\subset\;
   \hbox{series-parallel graphs}   \;=\;  {\cal G}_2^<
\ee
and we have shown that the chromatic roots of generalized theta graphs
$\Theta^{(s,p)}$ are unbounded.
(In fact, the family $\Theta^{(2,p)} \simeq K_{2,p}$
 already has unbounded chromatic roots:  see \cite{gen_theta}.)

\medskip

2) Read and Royle \cite[pp.~1027--1028]{Read_91} have observed empirically
that for cubic graphs with a fixed number of vertices,
the graphs with high girth tend to contribute the chromatic roots
with smallest real part.  It would be valuable to obtain a better
understanding of this phenomenon.  A first conjecture might be that
there is a universal lower bound on the real part of a chromatic root
in terms of the girth, but this is false:
the graphs $\Theta^{(3,p)}$ all have girth = circumference = 6,
but their chromatic roots accumulate as $p \to\infty$ on the hyperbola
\be
   (\imag q)^2 \,-\, (\real q - \smfrac{3}{2})^2   \;=\;   \smfrac{1}{4}
\ee
[set $s=3$ and $v=-1$ in \reff{fixed_s_curve}],
which contains points with arbitrarily negative real part.

\medskip

3) The construction employed in this paper is based on concatenating
2-rooted subgraphs (here single edges, but that can be generalized)
in order to create a larger 2-rooted subgraph with some desired value
of the effective coupling $v_{\hboxscript{eff}}$.
The situation may thus be very different for 3-connected graphs,
in which 2-rooted subgraphs of more than a single edge cannot occur.
It is thus reasonable to ask:
What is the closure of the set of chromatic roots of
{\em 3-connected}\/ graphs?
The answer, as Bill Jackson has pointed out to me,
is ``still the whole complex plane''.
To see this, consider the graphs $\Theta^{(s,p)} + K_n$
for fixed $n$ and varying $s,p$:
they are $(n+2)$-connected,
and their chromatic roots (taken together)
are dense in the region $|q - (n+1)| \ge 1$.
In particular, considering both $n$ and $n+2$,
the chromatic roots are dense in the whole complex plane.
So even arbitrarily high connectedness does not, by itself,
stop the chromatic roots from being dense in the whole complex plane.

On the other hand, the graphs $\Theta^{(s,p)} + K_n$ are non-planar
whenever $n \ge 1$ and $p \ge 3$.
This suggests posing a more restricted question:
\begin{question}
What is the closure of the set of chromatic roots for 3-connected
 {\em planar}\/ graphs?
\end{question}
Here the answer may well be much smaller than
$\C$ or $\C \setminus \{|q-1| < 1\}$.
But it will certainly not be a bounded set,
for as Bill Jackson has pointed out to me,
the bipyramids $C_n + \bar{K}_2$ are 4-connected plane triangulations,
but their chromatic roots are unbounded \cite{Read_88,Shrock_97a,Shrock_97c}.
%% Compare to Jackson \cite{Jackson_93} and Thomassen \cite{Thomassen_97}'s
%%   comments regarding the possible significance of 3-connectedness
%%   for {\em real}\/ roots.

\medskip

4) The chromatic roots of generalized theta graphs
have been further studied in \cite{Shrock_98e,Brown-Hickman_99a,gen_theta}.
In particular, Shrock and Tsai \cite[Section 3]{Shrock_98e}
plot the limiting set \reff{fixed_s_curve} of chromatic roots
as $p\to\infty$ for fixed $s=3,4,5$.
Brown, Hickman, Sokal and Wagner \cite{gen_theta}
prove that the chromatic roots of $\Theta_{s_1,\ldots,s_p}$
are bounded in magnitude by $[1 + o(1)] \, p/\log p$,
uniformly in the path lengths $s_1,\ldots,s_p$;
moreover, they prove that for $\Theta^{(2,p)} \simeq K_{2,p}$
this bound is sharp.

%% {\bf ADD CONJECTURE:}  For a graph of maximum degree $\Delta$,
%% all the real chromatic roots lie in the interval $[0,\Delta]$.
%% Indeed, we conjecture that $P_G(q)/q$ is strongly nonnegative at $q=\Delta$
%% (see Definition~\ref{def_strongly_positive}).

\section*{Acknowledgments}
I wish to thank Jason Brown, Roberto Fern\'andez, Walter Hayman, Bill Jackson,
Henry McKean, Criel Merino, Robert Shrock, Dave Wagner, Norman Weiss,
Dominic Welsh and Douglas Woodall
for valuable conversations and/or correspondence.
I especially wish to thank Norman Weiss for helpful comments
on a previous draft of this manuscript.
Finally, I wish to thank Herbert Spohn for suggesting, circa 1993,
that I study Potts models on hierarchical versions
of the graphs $\Theta^{(s,p)}$
\cite{Griffiths_82,Derrida_83,Derrida_84,Itzykson_85,Peitgen_85a,Peitgen_85b,%
Peitgen_86a,Peitgen_86b,Yung_91,Bleher_91};
though that work is still in progress, it directly inspired the present paper.

I also wish to thank an anonymous referee for comments on the
first version of this paper that led to improvements (I hope)
in the exposition.

This research was supported in part by
U.S.\ National Science Foundation grants PHY--9520978, PHY--9900769
and PHY--0099393.
Some of the work took place
during a Visiting Fellowship at All Souls College, Oxford,
where it was supported in part by Engineering and Physical Sciences
Research Council grant GR/M 71626
and aided by the warm hospitality of John Cardy and the
Department of Theoretical Physics.

\appendix
\section{Simple Proof of the Brown--Hickman Theorem on Chromatic Roots
   of Large Subdivisions}
  \label{app.A}

In this appendix I use the formula \reff{eq2.series}/\reff{eq2.seriesAB}
for series reduction of Potts edges to give a simple proof of the
Brown--Hickman \cite{Brown-Hickman_99b} theorem on chromatic roots
of large subdivisions.  This proof gives a further illustration
of the utility of considering the general Potts-model partition function
$Z_G(\qve)$ with not-necessarily-equal edge weights $v_e$,
even if one is ultimately interested
in the dichromatic polynomial $Z_G(q,v)$ or the chromatic polynomial $P_G(q)$.

\begin{lemma}
  \label{lemmaA.1}
Let $G$ be a graph of $n$ vertices and $m$ edges,
and let all the edge weights satisfy $|v_e| \le \delta$.
Then all the roots of $Z_G(\qve)$ lie in the disc
$|q| \le \max(C, C^{1/(n-1)})$ where $C = (1+\delta)^m - 1$.
\end{lemma}

\proof
{}From \reff{eq1.2} we have $Z_G(\qve) = \sum_{k=1}^n a_k q^k$
with $a_n = 1$ and
$\sum_{k=1}^{n-1} |a_k| \le (1+\delta)^m - 1 = C$.
Then
\be
   |Z_G(\qve)|  \;\ge\;  |q|^n  \,-\,  \sum\limits_{k=1}^{n-1} |a_k| \, |q|^k
     \;\ge\;  |q|^n  \,-\, \left. \cases{ C |q|^{n-1}  & if $|q| \ge 1$  \cr
                                          \noalign{\vskip 2mm}
                                          C |q|        & if $|q| \le 1$  \cr
                                        }
                           \right\}
   \;.
\ee
So $Z_G \neq 0$ if $|q| > \max(C, C^{1/(n-1)})$.
\qed

\noindent
{\bf Remark.}
This bound is far from being quantitatively optimal;
much better bounds, which depend on the maximum degree $\Delta$
but not on $n$ or $m$, can be obtained using the methods of \cite{Sokal_00}.
But Lemma~\ref{lemmaA.1} is good enough for our present purposes,
as we do not seek any uniformity in $G$.

\bigskip

Given a finite graph $G=(V,E)$ and a family of integers
${\bf k} = \{k_e\}_{e \in E} \ge 1$,
we define $G^{\bf k}$ to be the graph in which the edge $e$ of $G$
is subdivided into $k_e$ edges in series.

\begin{theorem}
   \label{thmA.2}
Let $G$ be a finite graph, let $v$ be a complex number,
and let $\epsilon > 0$.
Then there exists $K < \infty$
such that $k_e \ge K$ for all edges $e$ implies that
all the roots of $Z_{G^{\bf k}}(q,v)$ lie in the disc
$|q+v| \le (1+\epsilon) |v|$.
\end{theorem}

\proof
Let $G$ have $n$ vertices and $m$ edges.
The claim is trivial if $v=0$, so assume that $v \neq 0$.
By the series reduction formula \reff{eq2.series}/\reff{eq2.seriesAB},
we have
\be
   Z_{G^{\bf k}}(q,v)  \;=\;
   \left( \prod\limits_{e \in E} {(q+v)^{k_e} - v^{k_e} \over q} \right)
   \times\,
   Z_G(q, \{v_{\hboxscript{e,eff}}\})
 \label{eqA.star}
\ee
where
\be
   v_{\hboxscript{e,eff}}   \;=\;
   {q \, v^{k_e}  \over  (q+v)^{k_e} - v^{k_e}}
   \;.
\ee
The zeros of the prefactor in \reff{eqA.star}
all lie on the circle $|q+v| = |v|$.
So let us assume that $|(q+v)/v| = R > 1+\epsilon$;
we shall show that $Z_G(q, \{v_{\hboxscript{e,eff}}\}) \neq 0$
provided that $K$ is chosen sufficiently large
(depending only on $\epsilon$, $n$, $m$ and $|v|$).
We have $\epsilon < R-1 \le |q/v| \le R+1$ and hence
\be
   |v_{\hboxscript{e,eff}}|  \;\le\;  {R+1 \over R^K -1} \, |v|
      \;<\;  {2 + \epsilon  \over (1+\epsilon)^K - 1} \, |v|
      \;\equiv\; \delta
\ee
for all edges $e$ [since $(R+1)/(R^K-1)$ is a decreasing function of $R$
for $R > 1$].
It follows from Lemma~\ref{lemmaA.1} that
$Z_G(q, \{v_{\hboxscript{e,eff}}\}) \neq 0$
provided that we choose $K$ large enough so that
$\delta \le (1+\alpha)^{1/m} - 1$
where $\alpha = \min[\epsilon |v|, (\epsilon |v|)^{1/(n-1)}]$.
\qed

\noindent
The Brown--Hickman theorem is the special case $v=-1$.

\section{Chromatic Roots of 2-Degenerate Graphs}
\label{app.B} 

Let us recall that the {\em degeneracy number}\/ of a graph $G$
is defined by
\be
   D(G)  \;=\;  \max\limits_{H \subseteq G} \delta(H)   \;,
\ee
where the max runs over all subgraphs $H$ of $G$,
and $\delta(H)$ is the minimum degree of $H$.
A graph $G$ is said to be {\em $k$-degenerate}\/ if $D(G) \le k$.
Equivalently, $G$ is $k$-degenerate if there exists
an ordering $x_1,x_2,\ldots,x_n$ of the vertex set of $G$
such that
\be
   \deg\nolimits_{G \restrict \{x_1,x_2,\ldots,x_j\}}(x_j)   \;\le\;  k
\ee
for $1 \le j \le n$,
where $G \restrict \{x_1,x_2,\ldots,x_j\}$ denotes the induced subgraph.

It is an elementary fact \cite[Theorem V.1]{Bollobas_98}
that all of the {\em integer}\/ chromatic roots of $G$
lie in the interval $[0,\, D(G)]$,
i.e.\ the chromatic number $\chi(G)$ is at most $D(G) + 1$.
On the other hand, Thomassen \cite[Theorem 3.9]{Thomassen_97}
has shown that the {\em real}\/ chromatic roots
cannot be bounded in terms of the degeneracy number:
indeed, there are 2-degenerate graphs with arbitrarily large
real chromatic roots.
Here we shall use Proposition~\ref{prop2.2} and equation \reff{eq2.thetaAB}
to give a simple proof of the formula underlying
(a slight generalization of) Thomassen's construction,
and to exhibit its exact range of applicability.

One sees immediately from \reff{veff_def}/\reff{eq2.thetaAB}
that the 2-rooted generalized theta graph $\Theta^{(s,p)}$
(rooted at the endvertices) has an effective coupling
\be
   \veff(q)   \;=\;
   \left( {C_s(q) \over A_s(q)} \right) ^{\! p}  \,-\, 1
 \label{veff_thetasp}
\ee
where
\begin{eqnarray}
   A_s(q)   & = &  {(q-1)^s - (-1)^s  \over q}             \\[4mm]
   C_s(q)   & = &  {(q-1)^s + (q-1)(-1)^s   \over q}
\end{eqnarray}
%
% so that in particular
% \begin{subeqnarray}
%    C_2(q)   & = &   q-1         \\[2mm]
%    C_3(q)   & = &   (q-1)(q-2)  \\[2mm]
%    C_4(q)   & = &   (q-1)(q^2 - 3q + 3)
% \end{subeqnarray}
%
So let $Q_s \subset \C$ be the set on which $|1 + \veff(q)| > 1$, i.e.
\be
   Q_s   \;=\;  \{ q \in \C \colon\;
                   |(q-1)^s + (q-1)(-1)^s| \,>\, |(q-1)^s - (-1)^s|
                \}
   \;,
\ee
so that in particular
\begin{subeqnarray}
   Q_2   & = &   \{ q \colon\;  \Re q \,>\, \smfrac{3}{2} \}   \\[2mm]
   Q_3   & = &   \{ q \colon\;  (\Im q)^2 \,-\, (\Re q - \smfrac{3}{2})^2
                                    \,>\, \smfrac{1}{4}   \}
\end{subeqnarray}
We then have the following easy extension of Thomassen's result:

\begin{theorem}
 \label{thmB.1}
Let $G$ be a loopless graph, and let $s$ be an integer $\ge 2$.
Then there exist 2-degenerate (hence 3-colorable) graphs
$\{ G'(p) \} _{p=1}^\infty$ and an integer $k \ge 0$ such that
\begin{itemize}
   \item[(a)]  $\lim\limits_{p \to\infty} C_s(q)^{-kp} P_{G'(p)}(q) = P_G(q)$
     uniformly on compact subsets of $Q_s$.
   \item[(b)]  If $G$ is connected, then so are the $G'(p)$.
   \item[(c)]  If $s$ is even and $G$ is bipartite, then so are the $G'(p)$.
   \item[(d)]  If $G$ is embeddable on a surface $S$, then so are the $G'(p)$.
\end{itemize}
In particular:
\begin{itemize}
   \item[(i)]  If $q_0 \in Q_s$ is a zero of $P_G$,
      then for all $\epsilon > 0$ and all sufficiently large $p$
      (depending on $\epsilon$) $P_{G'(p)}$ has a zero in
      the disc $|q-q_0| < \epsilon$.
   \item[(ii)] If $q_0 \in Q_s \cap \R$ is a zero of $P_G$ of odd multiplicity,
      then for all $\epsilon > 0$ and all sufficiently large $p$
      (depending on $\epsilon$) $P_{G'(p)}$ has a zero in
      the interval $(q_0 - \epsilon, q_0 + \epsilon)$.
\end{itemize}
\end{theorem}

\proof
We can assume without loss of generality that $G=(V,E)$ has no multiple edges
(this is just to simplify the notation).
For each $x \in V$, define
\be
   V'_x   \;=\;   \cases{ \{x\}   & if $x$ is an isolated vertex of $G$  \cr
                          \noalign{\vskip 4pt}
                          \{ (x,y) \colon\;  xy \in E \}  & otherwise    \cr
                        }
\ee
and let $V' = \bigcup_{x \in V} V'_x$.
Let $E' = \{ (x,y)(y,x) \colon\,  xy \in E \}$,
and let $E''$ be a set of edges (with vertex set $V'$)
chosen so that the vertex sets of the
connected components of the graph $(V',E'')$
are precisely the sets $V'_x$.
[For example, we could connect the vertices of each $V'_x$
 by a path or a cycle or a complete graph.]
Now let $G' = (V', E' \cup E'')$,
and let $Z(q,v)$ be the Potts-model partition function $Z_{G'}(q, \{v_e\})$
with weights $v_e = -1$ for $e \in E'$ and $v_e = v$ for $e \in E''$.
It then follows immediately from \reff{eq1.2} that
\be
   Z(q,v)   \;=\;  \sum\limits_{k=0}^{|E''|} v^k \, p_k(q)
\ee
with highest-order coefficient
\be
   p_{|E''|}(q)   \;=\;   P_G(q)
   \;.
\ee

Now let $G'(p)$ be obtained from $G'$
by replacing each edge in $E''$ by the 2-rooted graph $\Theta^{(s,p)}$
(rooted at the endvertices).
By construction, $G'(p)$ is 2-degenerate, hence 3-colorable.
Moreover, $G'(p)$ is connected if $G$ is,
and $G'(p)$ is bipartite if $G$ is and $s$ is even.
Finally, if $G$ is embedded on a surface $S$,
then $G'(p)$ can also be taken to be embedded on $S$,
by drawing the vertex $(x,y) \in V'_x$ on the edge $xy$ very near to $x$,
and by choosing $E''$ to be a path (or cycle)
connecting the vertices of each $V'_x$ in cyclic order as one circles $x$.

It follows immediately from
Proposition~\ref{prop2.2} and equation \reff{eq2.thetaAB} that
\be
   P_{G'(p)}(q)   \;=\;  A_s(q) ^{p |E''|} \, Z(q, \veff(q))
\ee
where $\veff(q)$ is given by \reff{veff_thetasp}.
In particular, for $q \in Q_s$ we have
$\veff(q) \to \infty$ as $p \to\infty$, so that
\be
   \lim\limits_{p \to\infty}  C_s(q)^{-p|E''|} P_{G'(p)}(q)   \;=\;  P_G(q)
   \;.
 \label{eq.B.identity}
\ee
Moreover, the convergence is clearly uniform on compact subsets of $Q_s$.

Claim (i) then follows immediately from (a) using Hurwitz' theorem
\cite[p.~262]{Remmert_91} 
and the fact that $P_G$ is not identically zero.
To prove claim (ii), choose $\epsilon > 0$ small enough so that
$P_G(q_0 - \epsilon)$ and $P_G(q_0 + \epsilon)$ have opposite signs,
and choose $p_0$ large enough so that
$C_s(q_0 \pm \epsilon)^{-p|E''|} P_{G'(p)}(q_0 \pm \epsilon)$
has the same sign as $P_G(q_0 \pm \epsilon)$
for all $p \ge p_0$;
then use the intermediate value theorem.
\qed

\medskip
\noindent
{\bf Remarks.}
1.  Thomassen \cite[Theorem 3.9]{Thomassen_97}
considers the special case $s = 2$ and
restricts attention to real $q > 2$.
He establishes \reff{eq.B.identity} for {\em integer}\/ $q > 2$
by counting colorings,
and then asserts that ``this also holds'' for real $q > 2$.
The assertion is correct but requires a bit of argument;
Proposition~\ref{prop2.2} provides a convenient way
of filling in the missing details,
and determining the exact set $Q_s$ on which \reff{eq.B.identity} holds.

2.  One consequence of this result is that the upper zero-free interval
for 3-colorable (or 2-degenerate) planar graphs is the same as that
for all planar graphs
(unless the largest chromatic roots have even multiplicity,
 which seems unlikely).
In commenting on this corollary, Thomassen \cite[p.~506]{Thomassen_97}
asserted that there exist planar graphs
with real chromatic roots arbitrarily close to 4;
but this assertion apparently arises from a misunderstanding of
the Beraha--Kahane \cite{Beraha_79} theorem that
$4_{\rm periodic} \times n_{\rm free}$ triangular lattices
have {\em complex}\/ chromatic roots arbitrarily close to 4.
In fact I do not know of any planar graphs with real chromatic roots
arbitrarily close to 4.
A study of triangular-lattice strips
\cite{Jacobsen-Salas-Sokal_transfer3}
has thus far found chromatic roots up to $\approx 3.64$;
and Baxter's work \cite{Baxter_87}
suggests that sufficiently wide and long pieces of the triangular lattice
will have real chromatic roots up to at least
the Beraha number $B_{14} = 4 \cos^2 (\pi/14) \approx 3.801938$
(see also \cite{Jacobsen-Salas-Sokal_transfer3}
 for further discussion).
Finally, Douglas Woodall (private communication)
has found a 23-vertex plane triangulation with a chromatic root at
$\approx 3.811475$.
%% letter_from_woodall.Apr_4_01
But I do not know of any planar graphs with chromatic roots
larger than this.
It is therefore conceivable that the upper zero-free interval for planar graphs
extends {\em below}\/ $q=4$.
I thank Carsten Thomassen and Douglas Woodall for correspondence
concerning these questions.

% \bigskip
% {\bf What can I say about $\bigcup\limits_{s=2}^\infty Q_s$???
%    Can I show that
%    $\bigcup\limits_i Q_{s_i} \supset \{q\colon\, |q-1| > 1\}$
%    for any increasing sequence $s_1 < s_2 < \ldots$???
%    In particular, what about $s_i = $ all integers or even integers???
%    This tells me the set on which I can assert that the closure of
%    the chromatic roots of 2-degenerate graphs equals that of all graphs.
%    But maybe that's not exciting, as we already know that for
%    $\Theta^{(s,p)}$ the roots are dense in $\{q\colon\, |q-1| > 1\}$!!!}

\end{document}